\documentclass{jfm}
\usepackage{}
\usepackage{graphicx,epsfig}
  \usepackage{amssymb}
  \usepackage{amsmath}
  \usepackage{mathrsfs}
  \usepackage{mathtools}
  \usepackage{tikz}
\usepackage{natbib}
\usepackage{newtxtext}
\usepackage{newtxmath}
  \usepackage{amsfonts}
  \usepackage{enumitem}
  \usepackage{placeins}
  \usepackage{lineno}
  \usepackage{psfrag} 
  \usepackage{hyperref}
\usepackage{color}
\usepackage[utf8]{inputenc}

\DeclareMathAlphabet\mathbfcal{OMS}{cmsy}{b}{n}

\hypersetup{
    colorlinks = true,
    urlcolor   = blue,
    citecolor  = black,
}
    
\DeclareRobustCommand\sampleline[1]{%
  \tikz\draw[#1] (0,0) (0,\the\dimexpr\fontdimen22\textfont2\relax)
  -- (2em,\the\dimexpr\fontdimen22\textfont2\relax);%
}

\ifCUPmtlplainloaded \else
  \checkfont{eurm10}
  \iffontfound
    \IfFileExists{upmath.sty}
      {\typeout{^^JFound AMS Euler Roman fonts on the system,
                   using the 'upmath' package.^^J}%
       \usepackage{upmath}}
      {\typeout{^^JFound AMS Euler Roman fonts on the system, but you
                   dont seem to have the}%
       \typeout{'upmath' package installed. JFM.cls can take advantage
                 of these fonts,^^Jif you use 'upmath' package.^^J}%
      }
  \else
  \fi
\fi

\ifCUPmtlplainloaded \else
  \checkfont{msam10}
  \iffontfound
    \IfFileExists{amssymb.sty}
      {\typeout{^^JFound AMS Symbol fonts on the system, using the
                'amssymb' package.^^J}%
       \usepackage{amssymb}%
         \let\leq=\leqslant
         \let\geq=\geqslant
      }{}
  \fi
\fi

\ifCUPmtlplainloaded \else
  \IfFileExists{amsbsy.sty}
    {\typeout{^^JFound the 'amsbsy' package on the system, using it.^^J}%
     \usepackage{amsbsy}}
    {}
\fi

\def\bnab{\mbox{\boldmath $ \nabla$}}
\newcommand{\bu}{\mathbf u}
\newcommand{\C}{\mathbf C}
\newcommand{\F}{\mathbf F}
\newcommand{\LL}{\mathbf L}
\newcommand{\G}{\mathbf G}
\newcommand{\cG}{\mathbfcal G}
\newcommand{\T}{\mathbf T}
\def\bvarphi{\mbox{\boldmath $ \varphi$}}
\def\bpsi{\mbox{\boldmath $ \psi$}}
\def\b_eta{\mbox{\boldmath $ \eta$}}

\title[]{Weakly nonlinear analysis of the viscoelastic instability in channel flow for finite and vanishing Reynolds numbers}

\author[G. Buza, J. Page and R. R. Kerswell]%
{Gergely Buza$^1$\thanks{gb643@cam.ac.uk}\ns
 Jacob Page$^2$\thanks{jacob.page@ed.ac.uk}\ns
and\ns Rich R. Kerswell$^1$\thanks{r.r.kerswell@damtp.cam.ac.uk}}

\affiliation{$^1$Department of Applied Mathematics and Theoretical Physics, University of Cambridge, CB3 0WA, UK.\\
$^2$School of Mathematics, University of Edinburgh, EH9 3FD, UK}

\date{?; revised ?; accepted ?.}

\begin{document}
\maketitle
\begin{abstract}
The recently-discovered centre-mode instability of rectilinear viscoelastic  shear flow (Garg et al. {\em Phy. Rev. Lett.} {\bf 121}, 024502, 2018) has offered an  explanation for the origin of elasto-inertial turbulence (EIT) which occurs at lower Weissenberg ($Wi$) numbers. 
In support of this, we show using weakly nonlinear analysis that the subcriticality found in Page et al. ({\em Phys. Rev. Lett.} {\bf 125}, 154501, 2020) is generic across the neutral curve with the instability only becoming supercritical at low Reynolds ($Re$) numbers and high $Wi$.  
We demonstrate that the instability can be viewed as purely elastic in origin even for $Re=O(10^3)$, rather than  `elasto-inertial', as the underlying shear does {\em not} energise the instability.  
It is also found that the introduction of a realistic maximum polymer extension length, $L_{max}$, in the FENE-P model moves the neutral curve closer to the inertialess $Re=0$ limit at a fixed ratio of solvent-to-solution viscosities, $\beta$. 
In the dilute limit ($\beta \rightarrow 1$) with $L_{max} =O(100)$, the linear instability can brought down to more physically-relevant $Wi\gtrsim 110$ at $\beta=0.98$, compared with the threshold $Wi=O(10^3)$ at $\beta=0.994$ reported recently by Khalid et al. (arXiv: 2103.06794) for  an Oldroyd-B fluid. 
Again the instability is subcritical implying that inertialess rectilinear viscoelastic shear flow is nonlinearly unstable - i.e. unstable to finite amplitude disturbances  - for even lower $Wi$.

\end{abstract}

\section{Introduction}
\label{sect:intro}

% general intro
Viscoelastic flows have been of interest ever since the observation 70 years ago that a substantial reduction in viscous drag on a wall of a pipe carrying turbulent flow is possible after adding only a few parts per millon of long-chain polymers \citep{Toms1948}.  Just as curiously, adding further polymer quickly saturates this effect when the so-called `maximum drag reduction' regime (MDR) is entered \citep{Virk1970}, with skin friction reduced by $\sim$$80$\% relative to its Newtonian value. 
Efforts to explain this phenomenon have naturally focussed on understanding how 
low polymer concentrations moderate Newtonian turbulence (NT) \citep[e.g.][]{Lumley1969,Tabor1986, Procaccia2008, White2008}. However, the discovery of a new form of viscoelastic turbulence - `elasto-inertial' turbulence (EIT) - in 2013 \citep{Samanta2013, Dubief2013, Sid2018}  which exists at large Reynolds number $Re = O(10^3)$ and Weissenberg number $Wi=O(10)$  has provided a competing and even less well understood possibility.  Provided $Wi$ is large enough, EIT can exist at much lower $Re$ than NT explaining what has been labelled in the past as `early turbulence' \citep{Jones1966,Goldstein1969, Hansen1974,  Draad1998, Samanta2013,Choueiri2018,Chandra2018}. At higher but fixed $Re$, it is also possible, as the polymer concentration is steadily increased from zero, to relaminarize NT before triggering EIT \citep{Choueiri2018,Chandra2018}. In DNS, increasing $Wi$ from a state of EIT quenches the flow down to a simple travelling wave  solution and presumably laminar flow if  $Wi$ is large enough  \citep[e.g. see figure 2 in][]{Page2020, Dubief2020}. At even higher $Re$, it is currently unclear whether  the two types of turbulence merge or co-exist, and how MDR fits into the situation remains an outstanding issue \citep[e.g.][]{xi2010, xi2012, graham2014,Samanta2013,Choueiri2018, Choueiri2021, Lopez2019}.

Further questions also exist as to how EIT relates to another form of viscoelastic turbulence - `elastic' turbulence (ET) - that was discovered a decade earlier \citep{Steinberg2000}. This is generated by the well-known `elastic'  linear instability of curved streamlines \citep{Larson1990, Shaqfeh1996} and exists at vanishingly small Reynolds numbers so inertial effects are unambiguously irrelevant for sustaining the turbulence. 
This elastic instability is also possible in planar geometries, but requires finite-amplitude disturbances to generate streamline curvature \citep{Meulenbroek2004,Morozov2007}. 
In contrast to the inertialess ET, a fairly large $Re$ is required for EIT, indicating that inertia is important here. 
This suggests that EIT and ET are distinct phenomena \citep[e.g. see figure 30][]{Chaudhary2021} yet they could still be two extremes of the same whole \citep{Samanta2013, Qin2019, Choueiri2021, Steinberg2021}. Finally, the underlying mechanism which sustains EIT has yet to be clarified \citep{Dubief2013, Terrapon2015, Sid2018, Shekar2018, Shekar2020, Page2020, Chaudhary2021}.

A major step forward in explaining the origin of EIT was made recently when a linear instability was found at relatively high $Wi \gtrsim 20$ which could reach down to a threshold $Re_c \approx 63$ in pipe flow \citep{Garg2018, Chaudhary2021}. This finding overturned a long held view that no new linear instability would appear by adding polymers to a Newtonian rectilinear shear flow \citep[see][for an extensive historical discussion of this point]{Chaudhary2019, Chaudhary2021}. This instability was also confirmed in channel flow \citep{Khalid2021a}  using an Oldroyd-B fluid but was found absent in an upper-convected Maxwell (UCM) fluid \citep{Chaudhary2019}. The instability is a centre-mode instability which has a phase speed close to the maximum base-flow speed and appears to need inertia  (finite $Re$) to exist: in a channel with an experimentally-relevant $\beta$ (the ratio of solvent-to-solution viscosities) of $0.9$ and elasticity number of $0.1$, the threshold $Re_c \approx 200$. %
%\GB{-either this or the $\beta$ value is wrong(?) and contradicts fig2-}\RK{I took this directly from the abstract of Khalid2021a and see it also needs the qualifier $E=0.1$ to make sense...thanks}
%
However, in the dilute limit ($\beta \rightarrow 1$) and in contrast with pipe flow,  \cite{Khalid2021a} found that $Re_c$ could be pushed down to $\approx 5$ by the time  $\beta$ reached $0.99$, albeit at very large $Wi$. Further computations \citep{Khalid2021b} have confirmed that the elastic limit of $Re=0$ can indeed be reached at $\beta=0.9905$ and $Wi\approx 2500$. 
Looking beyond the extreme value of $Wi$ --which {\em is} apparently achievable experimentally \citep{Steinberg2018,Shnapp2021}--
this result has established a fascinating connection between an instability which appears to need inertia, elasticity and solvent viscosity (finite $(1-\beta)$) and a purely elastic instability when $(1-\beta)$ is small enough (\cite{Khalid2021b} refer to this as an `ultra dilute' polymer solution).

However, EIT appears at lower $Wi$ \citep[figure 2 in][]{Page2020} and sometimes lower $Re$ at a given $Wi$ \citep[see figure 1b in][]{Choueiri2021} than the centre-mode instability. For example, in channel flow at $Re=1000$ and $\beta=0.9$ in a FENE-P fluid with $L_{max}=500$, EIT occurs around $Wi=20$, whereas the centre-mode instability threshold is $Wi \approx 70$ \citep[figure 2 (left) in][]{Page2020}. This means that if EIT is dynamically connected to this instability, the hierarchy of nonlinear solutions which emerge from the linear instability must be substantially subcritical, reaching to $Wi$ far below those of the neutral curve (and similarly for $Re$ for high enough $Wi$). This was confirmed in one specific case on the neutral curve - $(Re,Wi,\beta)=(60,26.9,0.9)$ -- where the bifurcation was  shown to be strongly subcritical with the branch of travelling waves solutions reaching down to $Wi=8.77$ \citep{Page2020}. Moreover, the travelling wave solutions adopt  a distinctive  `arrowhead' form in the polymer stress when $Wi$ is small enough which can be recognised as an intermittently-observed coherent structure in  the DNS of EIT \citep{Dubief2020}.  

% the purpose
The primary purpose of this paper is to  back this initial finding of subcriticality up by carrying out a systematic survey of whether the centre mode bifurcation is sub- or supercritical across the entire neutral curve for a typical value of $\beta$ of $0.9$ using weakly nonlinear analysis \citep{Stuart60, Watson60}.  In doing so, we also take the opportunity to confirm that the instability is present for a FENE-P fluid with reasonable maximum polymer extension $L_{max}$ (see \ref{eq:fenep_constitutive}) and, spurred on by the recent results of \citet{Khalid2021b}, explore how the presence of finite $L_{max}$ affects the dilute limit ($\beta \rightarrow 1$) where $Re=0$ can be reached. We also examine the energetic source term,  or terms, for the instability uncovering a consistent picture even on the part of the neutral curve reaching to high $Re$.

% the plan
The plan of the paper is as follows. In \S\ref{sect:fenep}, the FENE-P model is introduced and the presence or not of polymer diffusion as indicated by a  Schmidt number $Sc$ is discussed. The weakly nonlinear expansions are also introduced. While this is now an established method in the fluid dynamicists' toolbox, for viscoelastic models where the (coarse-grained) local polymer configuration is represented by a positive definite conformation tensor $\C$, there are some technicalities which need some attention. We follow the framework recently suggested by \cite{Hameduddin2018, Hameduddin2019} to treat this issue which requires a bit more formal development than is normal. Having set this up, \S\ref{sect:weaklynonlin} then presents the weakly nonlinear analysis which proceeds as usual albeit with a proxy for $\C$ being expanded instead of $\C$ itself.  Results in \S\ref{sect:results_wna} are arranged as follows: \S\ref{L=500} and \S\ref{sect:flowprediction}  consider $(\beta,L_{max})=(0.9,500)$ with $Sc\to\infty$; \S\ref{L=100} considers $(\beta,L_{max},Sc)=(0.9,100,10^6)$; \S\ref{sect:energy} performs an energy analysis over the neutral curves of \S\ref{L=500} and \ref{L=100}; and finally \S\ref{sect:degenbeta}  examines the $Re=0$ situation varying $\beta$ over the approximate range of $[0.97,0.99]$ for $Wi \leq 200$ and $L_{max} \in[40,100]$ ($Sc\to \infty$). Lastly, \S\ref{sect:conclusions} presents a discussion of the paper's results.

%--------------------
%
% 2
%
%--------------------
\section{Formulation}
\label{sect:fenep}

We consider pressure-driven viscoelastic flow between two parallel, stationary, rigid plates separated by a distance $2h$ and assume that the flow is  governed by the FENE-P model
\begin{subequations} \label{eq:fenep}
\begin{align}
    \partial_t \bu + \left( \bu \cdot \bnab \right) \bu + \bnab p &= \frac{\beta}{Re} \Delta \bu + \frac{(1-\beta)}{Re} \bnab \cdot \T(\C), \label{eq:fenep_u} \\
    \bnab \cdot \bu &=0, \label{eq:fenep_incomp}\\
    \partial_t \C + \left( \bu \cdot \bnab\right) \C + \T(\C) &= \C \cdot \bnab \bu  +  \left( \bnab \bu \right)^{T} \cdot \C + \frac{1}{Re Sc} \Delta \C. \label{eq:fenep_C}
 % this isnt consistent with prev paper, disagrees with fluid conventions
\end{align}
% C (t,x) kellene
The constitutive relation for the polymer stress, $\T$, is given by the Peterlin function
\begin{equation}
    \T(\C) := \frac{1}{Wi} \Big( f (\mathrm{tr} \, \C) \C - {\mathbf I} \Big), \quad {\rm where} \quad 
f(x) := \left(1-\frac{x-3}{L^2_{max}}\right)^{-1}
    \label{eq:fenep_constitutive}
\end{equation}
\end{subequations}
with $L_{max}$ denoting the maximum extensibility of polymer chains.
Here $\C \in \mathrm{Pos}(3)$ (the set of  positive definite 3x3 matrices) is the polymer conformation tensor and
$\beta \in [0,1]$ denotes the viscosity ratio, $\beta:= \nu_s/\nu$, where $\nu_s$ and $\nu_p=\nu-\nu_s$ are the solvent and polymer contributions to the total kinematic viscosity $\nu$. The equations are non-dimensionalised by $h$ and the bulk speed  
\begin{equation}
U_b:= \frac{1}{2h} \int^h_{-h} u_x\, dy
\end{equation}
which, through adjusting the pressure gradient appropriately, is kept constant so that the Reynolds and Weissenberg numbers are defined as
\begin{equation}
Re:= \frac{hU_b}{\nu}, \quad Wi:= \frac{\tau U_b}{h}
\end{equation}
where $\tau$ is the polymer relaxation time.  Polymer diffusion - the last term in Eq.~\eqref{eq:fenep_C}  - is often omitted as the typical magnitude of the Schmidt number, $Sc \sim O( 10^6)$. Here it is retained throughout the nonlinear analysis to: 1. allow a more realistic comparison with results from direct numerical simulations (DNS), where a relatively low Schmidt number ($Sc \sim O( 10^3)$) is required for the solver to converge \citep{Page2020}, 
%
%\RK{NB only worth saying this if we compare WNA with DNS...} \GB{-yes, but this might actually happen with the modifications - if not, just change the sentence from 'DNS' to 'branch contionuation'-}
% JP -- left the above comment for a revision or future paper. Can leave DNS here but ref literature  
and 2. assess its importance  more generally. 
Non-slip boundary conditions are imposed on the velocity field. If an infinite Schmidt number $Sc$ is considered, no boundary conditions for the conformation tensor $\C$ are needed. In the case of finite Schmidt numbers, we apply $Sc\to\infty$ at the boundary to retain this situation \cite{Sid2018}. 
%
%\RK{(Jacob: since $Sc=10^6$ in reality, there are physical b.c.s to be applied and these presumably should be no-flux of $\C$ across the boundary...mention? say why we don't use these? is it because the DNS doesn't?)}.
% JP -- Again -- leaving the above for now. Thoughts: (1) I am not sure what a realistic BC is for C, is it really no flux? I have to think about that but it doesn't seem obvious to me that this should be the case and (2) I believe that the DNS BC is motivated to minimally perturb the solver from the Sc=oo case 
%

In the course of this work, we compute neutral curves for the recently-discovered centre mode instability in a pipe \citep{Garg2018, Chaudhary2021} and a channel flow \citep{Khalid2021a, Khalid2021b}.  The marginally-stable eigenfunctions form the basis of a weakly nonlinear expansion in the amplitude of the bifurcating solution. 
The key objective here is to ascertain whether the bifurcation is supercritical or subcritical. Subcriticality would indicate that bifurcated solutions exist beyond the parameter domain of linear instability, thereby implying that the flow is nonlinearly unstable - i.e. unstable  to sufficiently large amplitude disturbances - in new, potentially more interesting parameter regimes.
A case in point is the very recent discovery that the centre mode instability still operates at $Re=0$ albeit at very high $Wi=O(1000)$ and ultra-dilute polymer solutions of $1-\beta=O(10^{-3})$ \citep{Khalid2021b}. While these extremes are on the margins of physical relevance, a strongly-subcritical  instability could still see its consequences in the form of finite amplitude solutions at vastly different $Wi$ and $\beta$. 

%
% 2.1
%
\subsection{Base state} 
\label{sect:Base_state}

\begin{figure}%[h]
\includegraphics[width=0.65\textwidth]{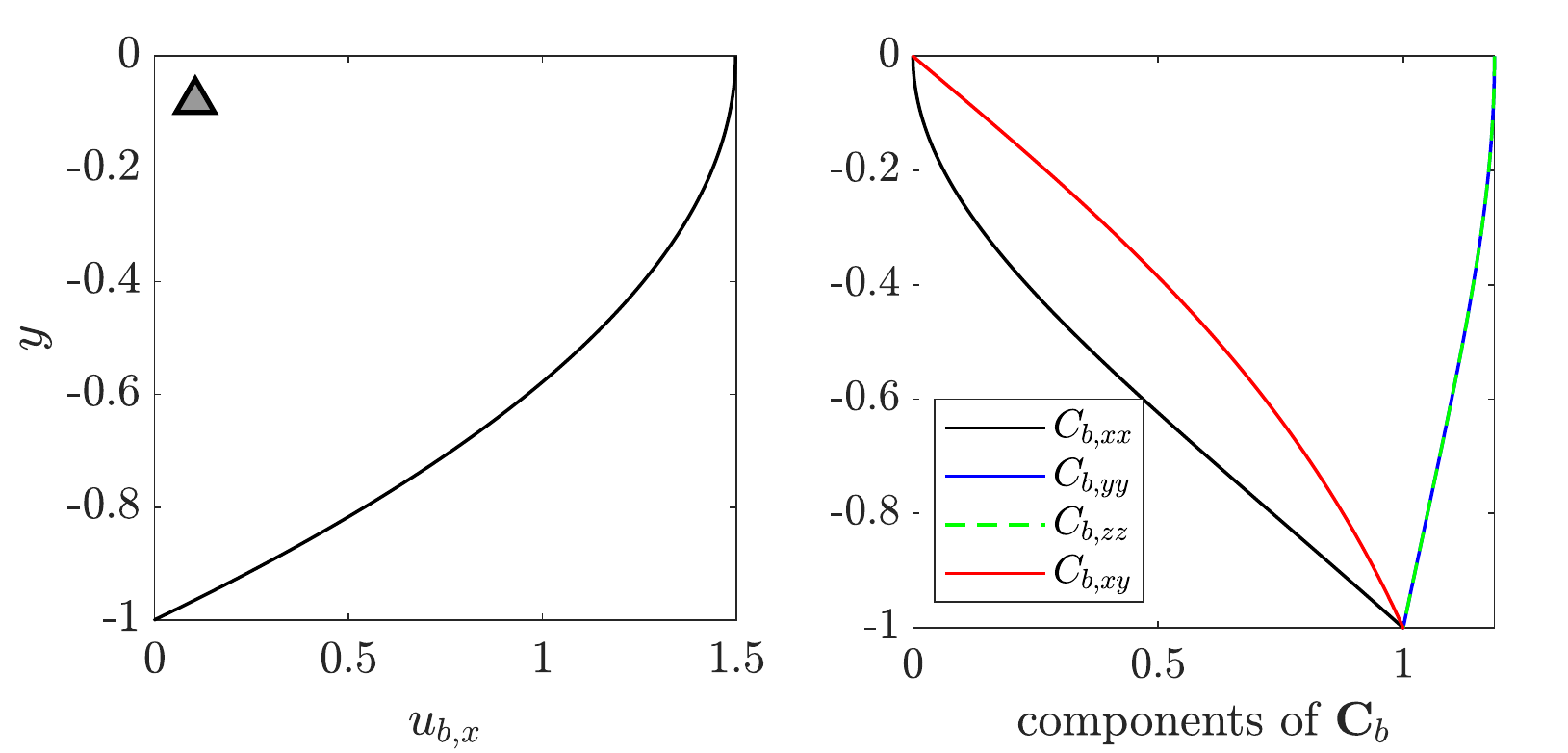}
\centering
\caption{Laminar base state at $\beta =0.9$, $L_{max} =500$, $Sc \to \infty$, $Wi = 60$, $Re = 68$.
The components of the base flow conformation tensor, $\C_b$, are normalized by their value at the bottom wall ($y = -1$). ($\left. C_{b,xx} \right\vert_{y=-1}=39235$, $\left. C_{b,yy} \right\vert_{y=-1}=\left. C_{b,zz} \right\vert_{y=-1}=0.84$, $\left. C_{b,xy} \right\vert_{y=-1} = 129$.)
}
\label{fig:base_state}
\end{figure}

The  base state to \eqref{eq:fenep_u}-\eqref{eq:fenep_C} is the steady unidirectional solution and satisfies the following reduced set of equations:
\begin{subequations} \label{eq:base_state}
\begin{align}
      &\partial_x p = \frac{\beta}{Re} \partial_{yy} u_x + \frac{(1-\beta)}{Re Wi}  \left[ \frac{\left(f (\mathrm{tr} \, \C)  \right)^2}{L^2_{max}} \mathrm{tr} (\partial_y C) C_{xy}+f (\mathrm{tr} \, \C) \partial_y C_{xy}\right], \label{eq:base_u} \\
     &\frac{1}{Wi} \left( f (\mathrm{tr} \, \C) C_{xx}-1 \right) = 2C_{xy} \partial_y u_x + \frac{1}{Re Sc} \partial_{yy} C_{xx}, \label{eq:base_Cxx}\\
     &\frac{1}{Wi} \left( f (\mathrm{tr} \, \C) C_{yy}-1 \right) =  \frac{1}{Re Sc} \partial_{yy} C_{yy}, \label{eq:base_Cyy}\\
     &\frac{1}{Wi} \left( f (\mathrm{tr} \, \C) C_{zz}-1 \right) =  \frac{1}{Re Sc} \partial_{yy} C_{zz}, \label{eq:base_Czz}\\
     &\frac{1}{Wi} \left( f (\mathrm{tr} \, \C) C_{xy} \right) = C_{yy} \partial_y u_x + \frac{1}{Re Sc} \partial_{yy} C_{xy}. \label{eq:base_Cxy}
\end{align}
\end{subequations}
where $\bu=u_x \hat{\bf x}+u_y \hat{\bf y}$. Since the $Re$ is based on the bulk speed, the applied pressure gradient is adjusted until the bulk speed is unity (after non-dimensionalisation)  \citep[e.g.][]{Samanta2013, Dubief2013,Sid2018, Dubief2020}.  Figure~\ref{fig:base_state} displays the base state  $(\bu_b,p_b,\C_b)$ for a particular parameter combination. It is worth remarking that $U_{max}$ is very nearly $1.5$ in units of $U_b$ for the parameter settings considered so that a $Wi$ based upon the bulk velocity (as here) is very close to two thirds of a Weissenberg number  based on $U_{max}$ \citep{Garg2018,Chaudhary2021,Khalid2021a,Khalid2021b}.
% \RK{is the base flow essentially parabolic so $U_b \approx 2/3U_{max}$? am conscious that Khalid et al. use $Wi$ based on $U_{max}$ whereas we use $Wi$ based on $U_b$: can we convert their $Wi=973$ (see later) to our $Wi$.}
%\GB{-it is very close to the $1.5(1-y^2)$ standard laminar profile from Newtonian turbulence, a bit distorted due to other eqs. I don't know what $U_b$ is supposed to mean here, but the difference between this and Khalid is that their base profile is $(1-y^2)$, making their volume flux $2/3$ over half the channel width as opposed to my $1$ - I can obviously convert their $Wi=973$ to mine (will be about 600) numerically, but I wonder if this is possible without computations -}

%
% Fig 1.
%

%
% 2.2
%
\subsection{Perturbative Expansions}
\label{sect:C_expansion}

The weakly nonlinear expansions for the velocity and pressure components are straightforwardly written in the form
\begin{equation}
    \bu = \bu_{b} + \sum_{k=1}^{N} \varepsilon^k \bu_{(k)}, \quad p = p_{b} + \sum_{k=1}^{N} \varepsilon^k p_{(k)}.
    \label{eq:u_expansion}
\end{equation}
However, the  conformation tensor, $\C$, calls for a more careful treatment, since the set of positive definite $3 \times 3$ matrices, $\mathrm{Pos} (3)$, cannot  be a vector space.
Instead, it may be endowed with the structure of a complete Riemannian manifold. Perturbations of order $\varepsilon^k$ still make sense in this setting, but one has to interpret the $\varepsilon^k$ distance in terms of the metric arising from the Riemannian structure of the manifold $\mathrm{Pos} (3)$. In developing perturbations for the conformation tensor, $\C$, we follow the framework of \cite{Hameduddin2018,Hameduddin2019} who focussed on precisely this issue. We may view $\C$ as the left Cauchy-Green tensor associated to the polymer deformation, i.e.,
$$
\C = \F \F^{T},
$$
where $\F$ denotes the deformation gradient with thermal equilibrium taken as the reference configuration.
A further decomposition of $\F$ into two successive deformations, which may be written as
\begin{equation}
    \F = \F_{b} \LL
    \label{eq:F_lamL}
\end{equation}
separates the deformation corresponding to the perturbation, $\LL$, from the deformation associated with the base state, which may be expressed as\footnote{This representation is not unique, any $\F_{b} = \C_{b}^{\frac{1}{2}} {\bf R}$ works with ${\bf R} \in \mathrm{SO}(3)$. The choice ${\bf R}={\bf I}$ is natural in the sense that it allows for a geodesic between $\C_{b}$ and $\C$ to be expressed solely in terms of $\F_{b}$ and $\G$.} 
$$
\F_{b} = \C_{b}^{\frac{1}{2}}.
$$
The fluctuating deformation gradient, $\LL$, has an associated left Cauchy-Green tensor $\G=\LL \LL^{T}$.
Combining these observations, we have that
\begin{equation}
    \C = \F_{b} \G \F_{b}^{T}.
\end{equation}
The tensor $\G$ is necessarily positive definite since $\C$ is, and by nature it acts as the conformation tensor representing the fluctuations of $\C$ around $\C_{b}$.
% ide a G eqt hameduddinbol

The evolution equation \eqref{eq:fenep_C} for the conformation tensor can be rewritten in terms of $\G$ as follows:
\begin{equation}
    \partial_t \G + (\bu \cdot \bnab) \G = 2 \mathrm{sym} \left(\G h(\bu) \right) - \F_{b}^{-1} \T \F_{b}^{-T},
    \label{eq:fenep_G}
\end{equation}
with
$$
h (\bu) = \F_{b}^{T} \cdot \bnab \bu \cdot \F_{b}^{-T}- \left(\F_{b}^{-1}  (\bu \cdot \bnab)\F_{b} \right)^{T}.
$$
% again, this disagrees with fluid conventions

As described by \cite{Hameduddin2019}, an additive expansion of the form \eqref{eq:u_expansion} no longer makes sense on $\mathrm{Pos}(3)$, since there is no a priori guarantee that the resulting $\C$ remains positive definite.
Instead, \cite{Hameduddin2019} proposed a multiplicative expansion based on the decomposition \eqref{eq:F_lamL} that consists of a series of successively smaller deformations, which may be written in the form
$$
\LL_{wnl} = \LL_{(1)}^{\varepsilon} \LL_{(2)}^{\varepsilon^2} \cdots  \LL_{(N)}^{\varepsilon^N}.
$$
The matrix $\LL_{wnl}$ may differ from $\LL$ given in \eqref{eq:F_lamL} by a rotation only.

Under the additional assumption that the $\LL_{(k)}^{\varepsilon^k}$ are rotation free with $\mathrm{det}(\LL_{(k)})>0$, each $\LL_{k}$ is positive definite.
The conformation tensors associated to these deformations are then given by $\G_{(k)}^{\varepsilon^k}= \LL_{(k)}^{\varepsilon^k} \left( \LL_{(k)}^{\varepsilon^k}\right)^{T}$.
To make sense of $\varepsilon$-magnitude perturbations, we make use of the Riemannian manifold structure of $\mathrm{Pos}(3)$.
In particular, the $\G_{(k)}^{\varepsilon^k}$ may be thought of as length $ \sim \vert \varepsilon \vert^k $ geodesics emanating from $\mathbf{I}$ on the manifold $\mathrm{Pos}(3)$.
That is, we may take $\cG_{(k)} \in T_\mathbf{I} \mathrm{Pos}(3) = \mathrm{Sym}(3)$ such that
$$
\G_{(k)}^{\varepsilon^k} = \mathrm{exp} \left(\varepsilon^k \cG_{(k)} \right),
$$
with
$$
d(\mathbf{I},\G_{(k)}^{\varepsilon^k}) = \vert \varepsilon \vert^k\Vert \cG_{(k)} \Vert_F,
$$
where $d$ is the metric induced by the Riemannian structure of $\mathrm{Pos}(3)$.
Note that this is analogous to weakly nonlinear expansions on vector spaces equipped with the Frobenius norm, only now we measure the corresponding distance on $\mathrm{Pos}(3)$ with the Riemannian metric.

This approach eventually leads to an expansion of the form
\begin{align}
    \G &= \mathrm{exp} \left(\varepsilon \frac{\cG_{(1)}}{2} \right) \cdots \mathrm{exp} \left(\varepsilon^{N-1} \frac{\cG_{(N-1)}}{2} \right) \mathrm{exp} \left(\varepsilon^{N} \cG_{(N)} \right) \mathrm{exp} \left(\varepsilon^{N-1} \frac{\cG_{(N-1)}}{2} \right) \cdots \mathrm{exp} \left(\varepsilon \frac{\cG_{(1)}}{2} \right) \nonumber \\
    &= \mathbf{I} + \varepsilon \cG_{(1)} + \varepsilon^2 \left( \cG_{(2)} + \frac{\cG_{(1)}^2}{2} \right) + \varepsilon^3 \left( \cG_{(3)} + \mathrm{sym} \left( \cG_{(1)} \cG_{(2)}\right) + \frac{\cG_{(3)}^3}{6}\right) + \ldots
    \label{eq:G_expansion}
\end{align}
% An expansion of this form is equivalent of a standard 
This representation of the weakly nonlinear terms is equivalent to a standard expansion for $\C$ of the form \eqref{eq:u_expansion}, as the operation
$ 
 \G_{(j)} \mapsto \F_{b} \G_{(j)} \F_{b}^{T}
$
serves as a bijection between the two solution sets, as long as $\F_{b} \in \mathrm{GL}(3)$.
Physically, this is always satisfied, as $\mathrm{det} (\F_{b}) = 0$ would imply that material elements are compressed to zero volume. % fix this

%In this sense, the new formulation provides no advantage over standard expansions.
While the new formulation does not in practice modify the mechanics of constructing a weakly nonlinear expansion,
the mathematical consistency of the approach yields a variety of tools for 
%However, due to the mathematical consistency of the approach, we gain access to a variety of tools that measure 
measuring perturbations on $\mathrm{Pos}(3)$ in the only suitable manner, according to the corresponding metric.
One such measure, which we shall use frequently in the sections to follow, is the geodesic distance from the mean, given by
\begin{equation}
 d(\C_b,\C) = d({\bf I},\G)= \sqrt{\mathrm{tr}  \cG^2}.
\label{distance}
\end{equation}
%

%------------------------------------
%
% 3
%
%------------------------------------
\section{Weakly nonlinear analysis}
\label{sect:weaklynonlin}

%In developing weakly nonlinear expansions, we restrict our analysis to the setting of channel flows.
Let
$
\bvarphi = (u_x,u_y,p,G_{xx},G_{yy},G_{zz},G_{xy})
$
denote the vector composed of all state variables.
This is further decomposed into two parts: a contribution from the base state and a fluctuating part as follows
%\footnote{The $\hat{(\, \cdot \,)}$ denotes the perturbations in all instances, i.e., $\hat{(\, \cdot \,)} = (\, \cdot \,) - (\, \cdot \,)_{b}$.}
$$
\bvarphi  = \bvarphi_{b} +\hat{\bvarphi},
$$
where the  interest is now  in solving the governing system \eqref{eq:fenep} for the perturbations $\hat{\bvarphi}$. The Peterlin function (\ref{eq:fenep_constitutive}) for $\T$ is first expanded
around the base conformation state, $\C_b$  as follows
\begin{equation}
\T(\C) = \T(\C_{b}) + D\T(\C_{b}) [ \hat{\C} ] + \frac12 D^2\T(\C_{b}) [ \hat{\C},\hat{\C} ] + \frac16 D^3\T(\C_{b}) [ \hat{\C},\hat{\C} ,\hat{\C}] + \ldots.
\label{eq:Texpansion}
\end{equation}
% with
% \begin{align*}
%     &DT(C_{b}) = \frac{1}{Wi} \left( \frac{(f(\mathrm{tr} \, C_{b}))^2}{L^2_{max}} C_{b,ij} \delta_{kl} +  f(\mathrm{tr} \, C_{b}) \delta_{ik} \delta_{jl}\right) e_i \otimes e_j \otimes e_k \otimes e_l, \\
%   &D^2T(C_{b})= \frac{1}{Wi} \left( \frac{2(f(\mathrm{tr} \, C_{b}))^3}{L^4_{max}} C_{b,ij} \delta_{kl} \delta_{mn} +  \frac{(f(\mathrm{tr} \, C_{b}))^2}{L^2_{max}} ( \delta_{ik} \delta_{jl} \delta_{mn} + \delta_{im} \delta_{jn} \delta_{kl})\right) e_i \otimes \cdots \otimes e_n, \\
%   &D^3T(C_{b}) = \frac{1}{Wi} \Bigg( \frac{6(f(\mathrm{tr} \, C_{b}))^4}{L^6_{max}} C_{b,ij} \delta_{kl} \delta_{mn} \delta_{op} \\ 
%     & \qquad \qquad \quad +  \frac{2(f(\mathrm{tr} \, C_{b}))^3}{L^4_{max}} ( \delta_{ik} \delta_{jl} \delta_{mn} \delta_{op} + \delta_{im} \delta_{jn} \delta_{kl} \delta_{op} + \delta_{io} \delta_{jp} \delta_{kl} \delta_{mn})\Bigg) e_i \otimes \cdots \otimes e_p.
% \end{align*}
% the above should be deleted from the actual paper.
For the analysis which follows, it suffices to perform the above expansion \eqref{eq:Texpansion} up to third order and to compress the notation, we shall only consider $Wi$ and $Re$ as varying parameters. The others, $\beta$ and $Sc$, are assumed fixed but similar expansions for them may be obtained in an analogous fashion.
After a subtraction of the laminar solution, equation \eqref{eq:fenep} can be written in an operator form locally around the base state $(\bu_b,\C_b)$ as
\begin{equation}
    \mathcal{L}\left(Re,Wi \right) \left[ \hat{\bvarphi} \right]
    +\mathcal{B}\left(Re,Wi \right) \left[ \hat{\bvarphi}, \hat{\bvarphi}\right]
    +\mathcal{T}\left(Re,Wi \right) \left[ \hat{\bvarphi},\hat{\bvarphi},\hat{\bvarphi} \right] ={\bf 0},
    \label{eq:mainoperatoreq}
\end{equation}
where $\mathcal{L}\left(Re,Wi \right)$ is linear, $\mathcal{B}\left(Re,Wi \right)$ is bilinear and $\mathcal{T}\left(Re,Wi \right) $ is symmetric trilinear.
These are given explicitly as
\begin{align*}
&\mathcal{L}\left(Re,Wi \right) \left[ \hat{\bvarphi} \right] = 
 \begin{pmatrix}
 \partial_t \hat{\bu} + (\bu_{b} \cdot \bnab) \hat{\bu} + ( \hat{\bu} \cdot \bnab)\bu_{b} + \bnab \hat{p} -\frac{\beta}{Re} \Delta \hat{\bu} - \frac{1-\beta}{Re} \bnab \cdot \left( D\T(\C_{b})\left[ \F_{b} \hat{\G} \F_{b}^{T} \right]\right)
\\
\\
 \bnab \cdot \hat{\bu} 
\\
\\
 \partial_t \hat{\G} + (\bu_{b} \cdot \bnab) \hat{\G} -2 \mathrm{sym} \left( h\left( \hat{\bu} \right) + \hat{\G} h(\bu_{b}) \right) +  \F_{b}^{-1} D\T(\C_{b})\left[ \F_{b} \hat{\G} \F_{b}^{T} \right] \F_{b}^{-T} \\
\hspace{6cm}- \frac{1}{ReSc} \F_{b}^{-1} \Delta \left( \F_{b} \hat{\G} \F_{b}^{T}\right)\F_{b}^{-T}
\\
\end{pmatrix}, 
\\
\\
&\mathcal{B}\left(Re,Wi \right) \left[ \hat{\bvarphi}_1,\hat{\bvarphi}_2 \right] =
\begin{pmatrix}
 (\hat{\bu}_1 \cdot \bnab) \hat{\bu}_2  - \frac{1-\beta}{2Re} \bnab \cdot \left( D^2\T(\C_{b})\left[ \F_{b} \hat{\G}_1 \F_{b}^{T}, \F_{b} \hat{\G}_2 \F_{b}^{T} \right]\right)\\ \\
0\\ \\
 (\hat{\bu}_1 \cdot \bnab) \hat{\G}_2 -2 \mathrm{sym} \left(  \hat{\G}_1 h(\hat{\bu}_2) \right) + \frac{1}{2} \F_{b}^{-1} D^2\T(\C_{b})\left[ \F_{b} \hat{\G}_1 \F_{b}^{T},\F_{b} \hat{\G}_2 \F_{b}^{T} \right] \F_{b}^{-T}
\end{pmatrix}, 
\\
\\
&\mathcal{T}\left(Re,Wi \right) \left[ \hat{\bvarphi}_1,\hat{\bvarphi}_2 ,\hat{\bvarphi}_3 \right] =
\begin{pmatrix}
- \frac{1-\beta}{6Re} \bnab \cdot \left( D^3\T(\C_{b})\left[ \F_{b} \hat{\G}_1 \F_{b}^{T}, \F_{b} \hat{\G}_2 \F_{b}^{T} ,\F_{b} \hat{\G}_3 \F_{b}^{T}\right]\right)\\ \\
0\\ \\
\frac{1}{6} \F_{b}^{-1} D^3\T(\C_{b})\left[ \F_{b} \hat{\G}_1 \F_{b}^{T},\F_{b} \hat{\G}_2 \F_{b}^{T},\F_{b} \hat{\G}_3 \F_{b}^{T} \right] \F_{b}^{-T}
\end{pmatrix}.
\end{align*}
It's worth remarking that the base state $(\bu_b,\C_b)$ in the above operators depends on all parameter values $(Wi,Re,\beta,Sc)$ through \eqref{eq:base_state}.
Linear stability theory is concerned with the eigenvalue problem arising from the linearized equations, $\mathcal{L}\left(Re,Wi \right) \left[ \hat{\bvarphi} \right] = {\bf 0}$. 
In practice, this is formally addressed by assuming a specific form of the disturbance, and solving
\begin{equation}
\mathcal{L} \left(Re,Wi \right) \left[ \bvarphi_{(1,1)}(y) \exp (ikx - i \omega t)  \right] = 0,
\label{eq:linearstability}
\end{equation}
for pairs $(\omega,\bvarphi_{(1,1)})$, where $\omega = \omega_r + i \omega_i$ is the {\em a priori} unknown complex frequency, $\bvarphi_{(1,1)}$ is the associated eigenmode and $k$ is the prespecified wave number.

% kicsit sketchy alább beletenni az om_r -t 
Assume now that a bifurcation occurs at a certain triple $(Wi_L,Re_L,k)$, i.e., there exists an eigenmode of \eqref{eq:linearstability} such that its associated eigenfrequency is real (subsequently denoted by $\omega_L = \omega_{L,r}$), which marks the state of marginal stability in the temporal sense.
We wish to uncover how the eigenfunction $\varphi_{(1,1)}$ evolves as we move slightly away from the bifurcation point. 
For this, consider small perturbations to all relevant parameters of the form
$$
 (Wi,Re,\omega_r) = (Wi_L,Re_L,\omega_{r,L}) + \varepsilon^2 (Wi_1,Re_1,\omega_{r,1}) + \ldots,
$$
and formally expand the operator $\mathcal{L}$ around $(Re_L,Wi_L)$ as
$$
\mathcal{L}(Re_L+\varepsilon^2 Re_1,Wi_L+ \varepsilon^2 Wi_1) = \mathcal{L}(Re_L,Wi_L) + \varepsilon^2 Re_1 \mathcal{L}'_{Re}(Re_L,Wi_L) + \varepsilon^2 Wi_1 \mathcal{L}'_{Wi}(Re_L,Wi_L).
$$
The subtle difference here from standard weakly nonlinear expansions lies in the fact that now the base state obtained from \eqref{eq:base_state} depends on the parameters $Wi$ and $Re$. 
To make this clear and explicit, we write
\begin{align*}
    &\mathcal{L}'_{Re}(Re_L,Wi_L) = \left. \frac{d}{dRe} \right\vert_{(Re_L,Wi_L)} \mathcal{L} =
    \left. \left( \frac{\partial}{\partial Re}  + \frac{\partial u_{b,i}}{\partial Re} \frac{\partial}{\partial u_{b,i}} + \frac{\partial F_{b,ij}}{\partial Re} \frac{\partial}{\partial F_{b,ij}} \right) \right\vert_{(Re_L,Wi_L)} \mathcal{L},\\
    &\mathcal{L}'_{Wi}(Re_L,Wi_L)= \left. \frac{d}{dWi} \right\vert_{(Re_L,Wi_L)} \mathcal{L} =
    \left. \left( \frac{\partial}{\partial Wi}  + \frac{\partial u_{b,i}}{\partial Wi} \frac{\partial}{\partial u_{b,i}} + \frac{\partial F_{b,ij}}{\partial Wi} \frac{\partial}{\partial F_{b,ij}} \right) \right\vert_{(Re_L,Wi_L)} \mathcal{L},
\end{align*}
with
$$
\frac{\partial \mathcal{L}}{\partial Re}(Re_L,Wi_L) [\hat{\bvarphi}] = 
\begin{pmatrix}
\frac{\beta}{Re_L^2} \Delta \hat{\bu} + \frac{1-\beta}{Re_L^2} \bnab \cdot \left( D\T(\C_{b}) \left[ \F_{b} \hat{\G} \F_{b}^{T} \right] \right) \\
0 \\
\frac{1}{Re_L^2Sc} \F_{b}^{-1} \Delta \left( \F_{b} \hat{\G} \F_{b}^{T}\right)\F_{b}^{-T}
\end{pmatrix},
$$
and
$$
\frac{\partial \mathcal{L}}{\partial Wi}(Re_L,Wi_L) [\hat{\bvarphi}] = 
\begin{pmatrix}
 \frac{1-\beta}{Re_L Wi_L} \bnab \cdot \left( D\T(\C_{b}) \left[ \F_{b} \hat{\G} \F_{b}^{T} \right] \right) \\
0 \\
-\frac{1}{Wi_L} \F_{b}^{-1} D\T(\C_{b})\left[ \F_{b} \hat{\G} \F_{b}^{T} \right] \F_{b}^{-T}
\end{pmatrix}.
$$
Due to the complexity of the laminar equations \eqref{eq:base_state}, the base flow's dependence on the parameters is sought numerically, i.e., the terms $\partial u_{b,i} / \partial Re$ and $\partial F_{b,ij} / \partial Re$ - and the corresponding terms in the $Wi$ direction - are computed via a finite difference scheme. 
We note here that alternatively one could also compute the entirety of $\mathcal{L}'_{Re}$ (and $\mathcal{L}'_{Wi}$) with a finite difference scheme.

% do something different, valahogy hidald
To explore how the $\bvarphi_{(1,1)}$ wave develops as these parameters change, we seek solutions of \eqref{eq:mainoperatoreq} as a weakly nonlinear expansion of the form
\begin{equation}
    \bvarphi (t,x,y)  = \bvarphi_{b}(y) + \sum_{l=1}^N \sum_{q \in J_l} \varepsilon^l \left(  \bvarphi_{(l,q)} + \tilde{\bvarphi}_{(l,q)} \right) (y) \, \mathrm{exp} \big( iq(kx-\omega_r t) \big)+ O( \varepsilon^{N+1} ), 
    \label{eq:phi_expansion}
\end{equation}
where $J_l = \{-l, -l+2, \ldots, l-2,l \}$,
and $\tilde{\bvarphi}_{(l,q)}$ is the term that represents the dependence of $O( \varepsilon^l )$ perturbations on the lower order $\cG_{(j)}$ terms in \eqref{eq:G_expansion}.
For instance, $\tilde{\bvarphi}_{(1,q)} = 0,$ $q \in \{-1,1 \}$, and 
$$
\tilde{\bvarphi}_{(2,2)} = \frac{1}{2} \left(0,0,0, \left(\mathcal{G}_{(1,1)}^2\right)_{xx}, \left(\mathcal{G}_{(1,1)}^2\right)_{yy}, \left(\mathcal{G}_{(1,1)}^2\right)_{zz}, \left(\mathcal{G}_{(1,1)}^2\right)_{xy} \right).
$$
To simplify the notation, let
$$
E_q : (t,x) \mapsto \mathrm{exp} \big( iq(kx-\omega_{r,L} t) \big),
$$
and
$$
\mathcal{L}_q[\bvarphi] := \mathcal{L} [\bvarphi E_q].
$$
Now, upon substituting the specific form of $\hat{\bvarphi}$ from Eq.~\eqref{eq:phi_expansion} into \eqref{eq:mainoperatoreq}, we obtain a hierarchy of problems as follows:
\begin{subequations} \label{eq:WNL}
\begin{align}
    &O(\varepsilon ) : & &\mathcal{L}_1[\bvarphi_{(1,1)}]={\bf 0}, \label{eq:wnl_1}\\
    & O(\varepsilon^2 ) :& &\mathcal{L}_0 [\bvarphi_{(2,0)} + \tilde{\bvarphi}_{(2,0)}] + \mathcal{B}[\bvarphi_{(1,1)}E_1,\bvarphi_{(1,-1)}E_{-1}] + \mathcal{B}[\bvarphi_{(1,-1)}E_{-1},\bvarphi_{(1,1)}E_1] = {\bf 0},\\
    & & &\mathcal{L}_2[\bvarphi_{(2,2)}+ \tilde{\bvarphi}_{(2,2)}] + \mathcal{B}[\bvarphi_{(1,1)}E_1,\bvarphi_{(1,1)}E_{1}] = {\bf 0},\\
   &O(\varepsilon^3 )  : & &\mathcal{L}_1 [\bvarphi_{(3,1)} + \tilde{\bvarphi}_{(3,1)}] +\mathcal{B}[\bvarphi_{(1,-1)}E_{-1},( \bvarphi_{(2,2)}+ \tilde{\bvarphi}_{(2,2)})E_2] \nonumber \\
    & & &+\mathcal{B}[( \bvarphi_{(2,2)}+ \tilde{\bvarphi}_{(2,2)})E_2,\bvarphi_{(1,-1)}E_{-1}]
     + \mathcal{B}[\bvarphi_{(1,1)}E_1,\bvarphi_{(2,0)}+ \tilde{\bvarphi}_{(2,0)}]   \nonumber \\
     & & &+ \mathcal{B}[\bvarphi_{(2,0)}+ \tilde{\bvarphi}_{(2,0)},\bvarphi_{(1,1)}E_1] + 3 \mathcal{T}[\bvarphi_{(1,1)}E_1,\bvarphi_{(1,1)}E_1,\bvarphi_{(1,-1)}E_{-1}] \nonumber \\
    & & &+ Re_1 \mathcal{L}'_{Re}[\bvarphi_{(1,1)}E_1] + Wi_1 \mathcal{L}'_{Wi}[\bvarphi_{(1,1)}E_1] - i  \omega_{r,1} \bvarphi_{(1,1)}  =:\mathcal{L}_1 [\bvarphi_{(3,1)}] +\b_eta= {\bf 0},\label{eq:wnl_end} \\
    & & & \qquad \vdots \nonumber
\end{align}
\end{subequations}
where $\b_eta$ is the known part of the last equation \eqref{eq:wnl_end}. One subtlety in solving the hierarchy of problems is maintaining the constancy of the volumetric flux. This boils down to introducing a constant correction to the pressure gradient, $\partial_x p_{(2,0)}$, to ensure  $\bvarphi_{(2,0)}$ has zero flux.  Provided that the bifurcation is of codimension one, equation \eqref{eq:wnl_1} (equivalent to the linear problem, \eqref{eq:linearstability}) has a non-unique solution of the form 
\begin{equation}
    A \frac{\bvarphi_{(1,1)}}{\Vert \bvarphi_{(1,1)} \Vert_{L^2 \left([-1,1]; \mathbb{C}^7 \right)}}, \qquad A \in \mathbb{C}.
    \label{eq:Aphi}
\end{equation}
The aim is to map out the possible values of the steady-state amplitude $A$ in the parameter space $(Wi,Re)$.
Once an eigenmode of the form \eqref{eq:Aphi} is pushed through equations \eqref{eq:wnl_1} to \eqref{eq:wnl_end}, an explicit solvability condition can be derived, as detailed in the following.

%
% 3.1  Solvability
%
\subsection{Solvability condition}

Let us view the functions $\bvarphi_{(i,j)}:[-1,1] \rightarrow \mathbb{C}^7 $ as elements of $L^2 \left( [-1,1]; \mathbb{C}^7 \right) $.
The inner product on $L^2 \left( [-1,1]; \mathbb{C}^7 \right) $ is given by\footnote{
In the following, we use an $L^2$ inner product on matrix valued functions as well. In this case, we simply identify the matrices with vectors in the canonical way (i.e., we replace the $\mathbb{C}^7$ inner product below the integral with a Frobenius one).}
$$
\langle \bvarphi, \bpsi \rangle_{L^2\left( [-1,1]; \mathbb{C}^7 \right)} = \int_{ [-1,1]} \langle \bvarphi (y), \bpsi(y) \rangle_{\mathbb{C}^7} \, dy.
$$
The linear problem \eqref{eq:wnl_1} implies that $\mathcal{L}_1$ has a nontrivial kernel. 
Therefore, the Fredholm alternative theorem (for elliptic PDEs) 
%\footnote{
% Technically, $\mathcal{L}_1$ is not a uniformly elliptic operator as it stands. 
% However, the addition of an arbitrarily small $\delta \Delta p$, $0 < \delta \ll 1$, term to the incompressibility condition circumvents this issue.}) 
implies the existence of a finite dimensional subspace of solutions to the adjoint homogeneous problem 
$$
\mathcal{L}^*_1 [\bpsi] = {\bf 0},
$$
subject to the appropriate boundary conditions (matching those of the original problem).
Moreover, the original equation \eqref{eq:wnl_end} has a solution, $\bvarphi_{(3,1)}$, if and only if 
\begin{equation} 
    \langle \b_eta , \bpsi \rangle_{L^2\left( [-1,1]; \mathbb{C}^7 \right)} = 0, \qquad \forall \bpsi \in \mathrm{ker} \, \mathcal{L}^*_1 \text{ satisfying the boundary conditions.}
    \label{eq:Fredholm_condition}
\end{equation}
Assuming that the bifurcation is of codimension one, we know that $\mathrm{dim} \left( \mathrm{ker}\, \mathcal{L}^*_1 \right) = 1$, so it suffices to check \eqref{eq:Fredholm_condition} for any $\bpsi_1 \in \mathrm{ker} \, \mathcal{L}^*_1$ that satisfies the boundary conditions.
With this procedure, we obtain the complex solvability condition
\begin{equation}
    a Re_1 + b Wi_1 + c \vert A \vert^2 + d \omega_{r,1} = 0,
    \label{eq:solvability}
\end{equation}
where
\begin{align*}
   & & a := &\left\langle \mathcal{L}'_{Re}[\bvarphi_{(1,1)}E_1], \bpsi_1 \right\rangle_{L^2\left( [-1,1]; \mathbb{C}^7 \right)} \\
   & & b := &\left\langle \mathcal{L}'_{Wi}[\bvarphi_{(1,1)}E_1], \bpsi_1 \right\rangle_{L^2\left( [-1,1]; \mathbb{C}^7 \right)} \\
   & & c := &\Big\langle  \mathcal{B}[\bvarphi_{(1,-1)}E_{-1},( \bvarphi_{(2,2)}+ \tilde{\bvarphi}_{(2,2)})E_2] +\mathcal{B}[( \bvarphi_{(2,2)}+ \tilde{\bvarphi}_{(2,2)})E_2,\bvarphi_{(1,-1)}E_{-1}] \\  
   & & &+\mathcal{B}[\bvarphi_{(1,1)}E_1,\bvarphi_{(2,0)}+ \tilde{\bvarphi}_{(2,0)}]   + \mathcal{B}[\bvarphi_{(2,0)}+ \tilde{\bvarphi}_{(2,0)},\bvarphi_{(1,1)}E_1]  \\
   & & & + 3 \mathcal{T}[\bvarphi_{(1,1)}E_1,\bvarphi_{(1,1)}E_1,\bvarphi_{(1,-1)}E_{-1}]
    ,\bpsi_1 \Big\rangle_{L^2\left( [-1,1]; \mathbb{C}^7 \right)}, \\
    & &d := & \left\langle - i \bvarphi_{(1,1)} ,\bpsi_1 \right\rangle_{L^2\left( [-1,1]; \mathbb{C}^7 \right)}.
\end{align*}
Equation~\eqref{eq:solvability} gives the desired relationship between the parameters $(Wi_1,Re_1)$ and the steady state amplitude $A$, which allows us to track how these finite amplitude states emerge from the bifurcation point.

%
% 4
%
\section{Results}
\label{sect:results_wna}

As indicated above, we are interested in uncovering the nature of the initial bifurcation associated to the centre-mode instability first identified by \cite{Garg2018} in pipe flow and, most relevantly for us, later by  \cite{Khalid2021a} in channel flow. This previous work assumed an Oldroyd-B fluid which allows infinite polymer extension i.e. $L_{max} \to \infty $ for the FENE-P model \eqref{eq:fenep_constitutive}. Given this, our objectives in what follows are two-fold.  On the one hand, we want to explore the effects of finite extensibility on the aforementioned instability. And on the other,  with the aid of the weakly nonlinear analysis, we aim to identify parameter regions where the instability persists beyond the neutral curve to lower $Wi$ in particular.

%
% 4.1
%
\subsection{$\beta=0.9$, $L_{max}=500$ \& $Sc \to \infty$ }
\label{L=500}

%
%  upper branch (comparison with Page et al. 2020)
%
In order to test the weakly nonlinear analysis, we begin by examining the parameter regime considered by \cite{Page2020} where $\beta = 0.9$ and $L_{max}=500$. Using $Sc=10^3$ to stabilise their time-stepping code, \cite{Page2020} observed substantial subcriticality at $(Re,Wi,k) =(60,26.9,2)$ on the upper branch of the neutral curve since they were able to continue the branch of solutions down to $Wi =8.77$.  Figure~\ref{fig:L500NC} shows the neutral curve at $\beta = 0.9$, $L_{max} = 500$ with $Sc \to \infty$: see appendix \ref{sect:numerics} for numerical details. The neutral curve is insensitive to the choice of $Sc$ on the scale of Figure~\ref{fig:L500NC} provided it is $\gg 10^2$. Alongside the neutral curve, we display the results of the weakly nonlinear analysis by plotting a curve corresponding to a finite (small) steady state amplitude $\vert A \vert$, as obtained from the solvability condition \eqref{eq:solvability}. The linear instability is a Hopf bifurcation and so the steady state solutions are travelling waves (in $x$) with phase speed $\omega_r/k$ and a constant amplitude which decreases to zero at the neutral curve.  This finite amplitude curve in Figure~\ref{fig:L500NC} clearly indicates subcriticality along the upper branch of the neutral curve. Proceeding down to the lower branch of the curve, the Hopf bifurcation switches to being supercritical for $Wi\gtrsim 40$ (the red dashed line crosses the black neutral curve).

%
% Fig 2
%
\begin{figure}%[h]
% decrease the size of the lines leading to subritical/supercritical in these plots
\includegraphics[width=1\textwidth]{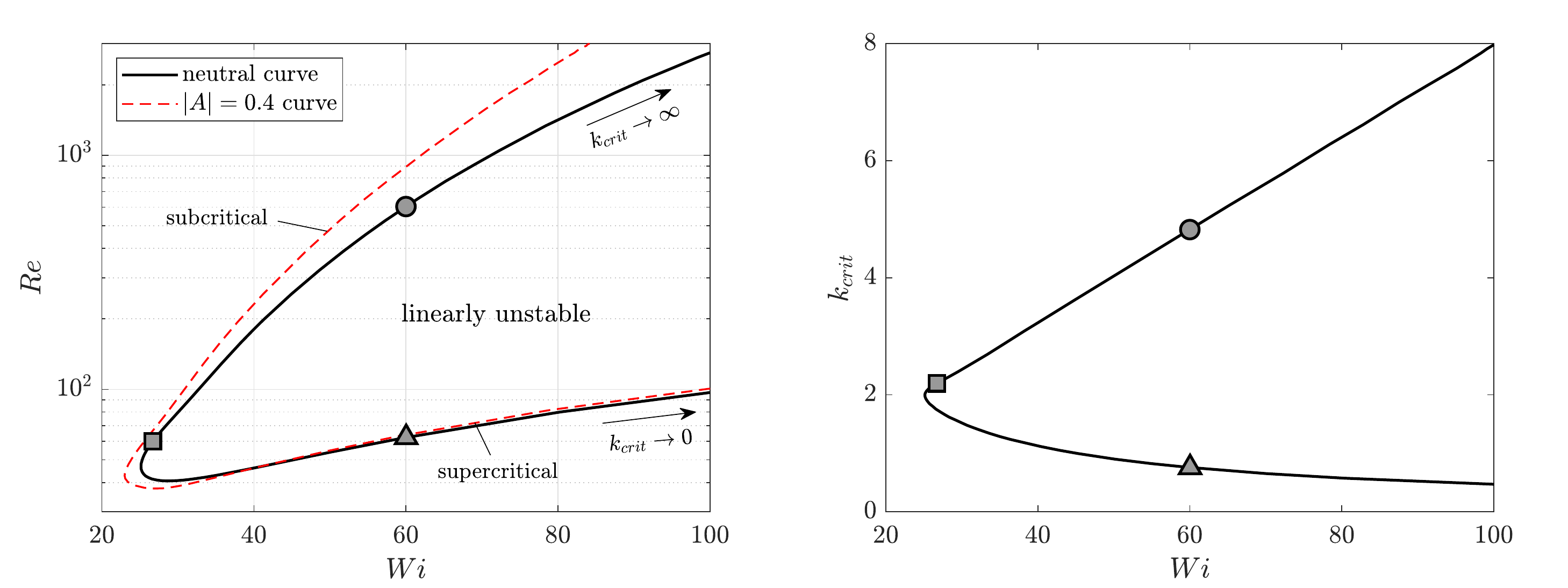}
\centering
\caption{(Left) Neutral curve corresponding to marginal linear stability at $\beta = 0.9$, $L_{max} = 500$, $Sc \to \infty$. Results of the weakly nonlinear analysis are shown in the form of a curve at steady state amplitude $\vert A \vert = 0.4$. (Right) The development of the critical wave number, $k_{crit}$, along the neutral curve. Since $k_{crit}$ varies monotonically along the neutral curve its provides a convenient  parametrization of it in subsequent figures.}
\label{fig:L500NC}
\end{figure}

%
% Fig 3
%
\begin{figure}%[h]
\includegraphics[width=1\textwidth]{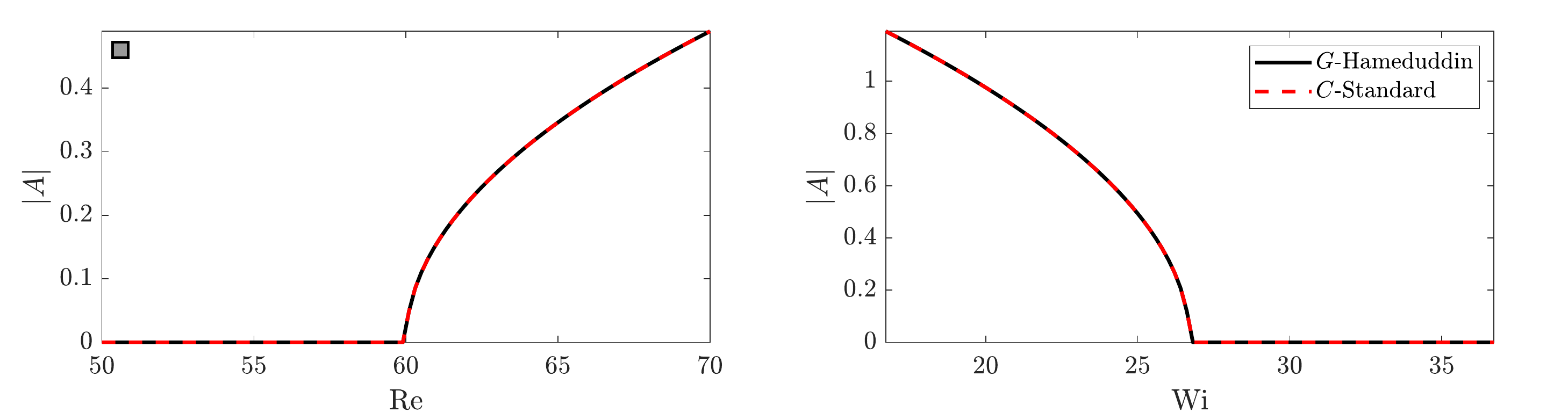}
\centering
\caption{Bifurcation diagrams at $(Wi,Re,k) \approx (27,60,2)$ ($\Box$). Only the unstable branch is displayed with a comparison of two methods for the expansion for the conformation tensor.}
\label{fig:Wi27Re60}
\end{figure}

Figure~\ref{fig:L500NC} confirms the subcritical behaviour observed by \cite{Page2020} at the point $(Wi,Re,k) \approx (27,60,2)$, which is marked by a shaded square $\Box$. The corresponding bifurcation diagrams with respect to model parameters $Wi$ and $Re$ are shown in Figure~\ref{fig:Wi27Re60}. In the figure, the newly developed approach for perturbative expansions of \cite{Hameduddin2019} (described in \S\ref{sect:C_expansion}) is compared with a standard expansion in the conformation tensor, $\C$. As mentioned in \S\ref{sect:C_expansion}, the two should be equivalent when considering quantities that only depend on $\C$. 
% otherwise one could argue that the choice of F*R matters, but still not really
In this context, the real advantage of using the form of expansions established in \S\ref{sect:C_expansion} is that we now have immediate access to quantities with tangible physical meaning (cf.\ Figures \ref{fig:EigenF11}, \ref{fig:EigenF20} and \ref{fig:EigenF22}).

\begin{figure}%[h]
\includegraphics[width=1\textwidth]{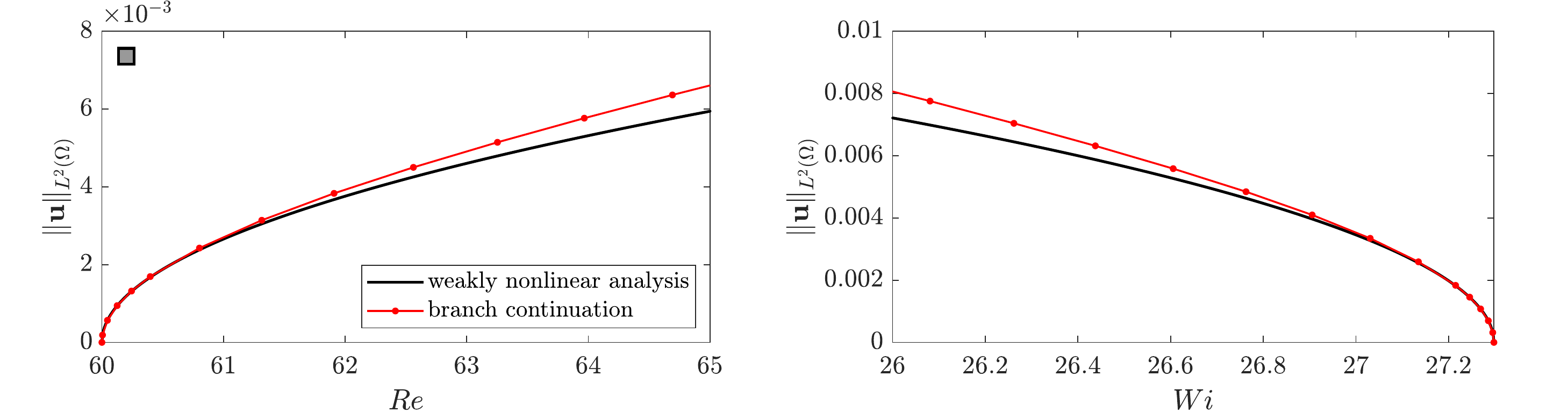}
\centering
\caption{Validation of the weakly nonlinear analysis at $(Wi,Re,k) \approx (27,60,2)$ ($\Box$) with a full branch continuation prediction. The $L^2$ norms are taken over the whole domain $\Omega = [0,2\pi/k]\times [-1,1]$.}
\label{fig:compSUBC}
\end{figure}

As a final check, results of the weakly nonlinear analysis are compared with a full  branch continuation computation (see appendix \ref{sect:numerics} for details of the method) in Figure~\ref{fig:compSUBC}. A finite but large Schmidt number of $Sc=10^3$ had to be selected for this comparison, as the latter method requires a diffusion term to produce reliable results. 
The curves are in good agreement - on top of each other near  the bifurcation point but then diverging slightly (not visible on the plots) as the amplitude increases (as they should).
This divergence, of course,  is because the weakly nonlinear analysis is based upon a  3-Fourier-mode expansion whereas the branch continuation curve is from a  40-mode Fourier expansion.
 
%
% Fig 4
%

\begin{figure}%[h]
\includegraphics[width=1\textwidth]{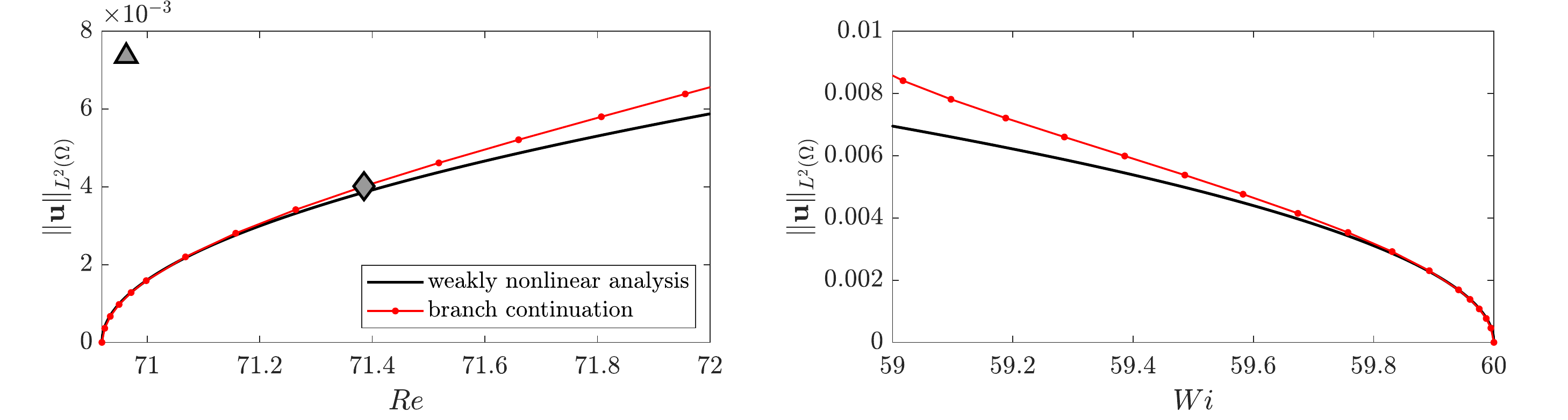}
\centering
\caption{Bifurcation diagrams at point $\triangle$. The $L^2$ norms are taken over the whole domain $\Omega = [0,2\pi/k]\times [-1,1]$.}
\label{fig:compSUPERC}
\end{figure}

\begin{figure}%[h]
% \psfrag{components of mathbfcalG11}[][][0.9]{components of $\cG_{(1,1)}$}
\includegraphics[width=1\textwidth]{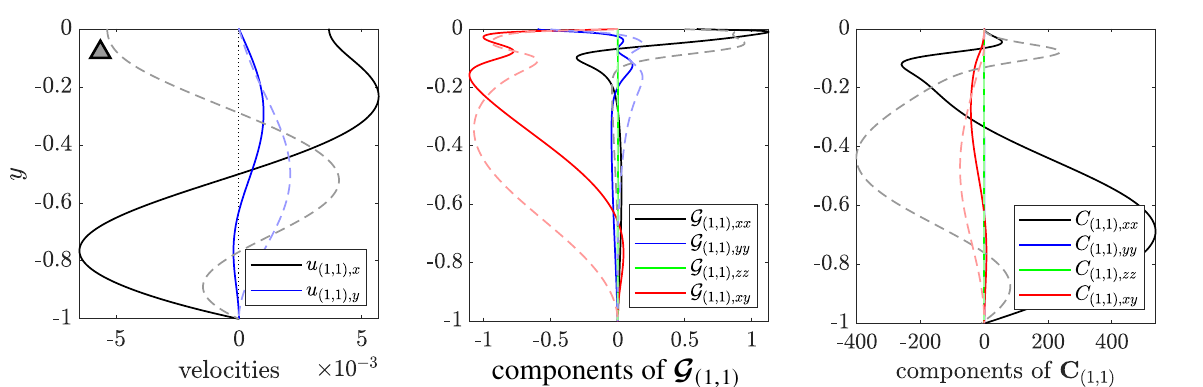}
\centering
\caption{Real (\sampleline{}) and imaginary (\sampleline{dashed}) parts of the unstable eigenfunction $\bvarphi_{(1,1)}$ at the point $\triangle$. (Left) Axial (streamwise) velocity $u_{(1,1),x}$ and vertical velocity $u_{(1,1),y}$. (Middle) All four nonzero components of $\cG_{(1,1)} \in T_\mathbf{I}\mathrm{Pos}(3)$, the tangent form of the polymer strain perturbation tensor. (Right) All four nonzero components of $\C_{(1,1)} \in \mathrm{Pos}(3)$, the corresponding fluctuation tensor from a standard expansion.}
\label{fig:EigenF11}
\end{figure}

\begin{figure}%[h]
% \psfrag{components of mathbfcalG20}[][][0.9]{components of $\cG_{(2,0)}$}
\includegraphics[width=\textwidth]{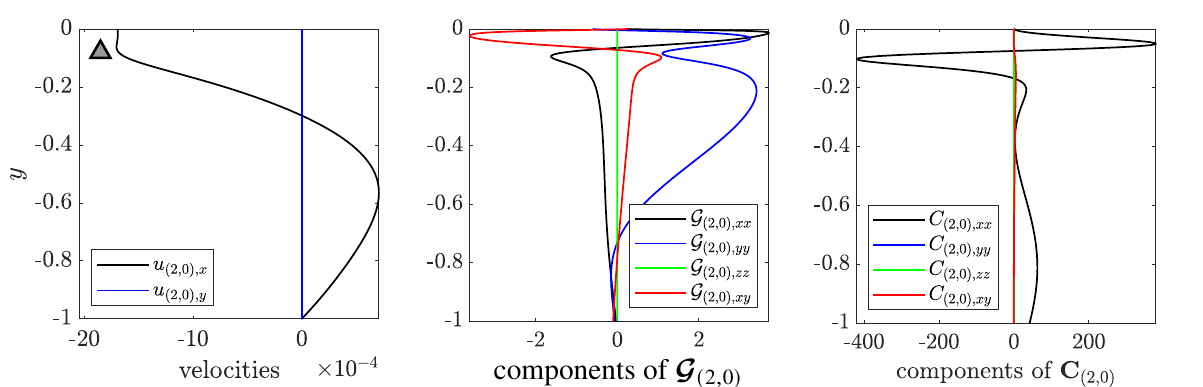}
\centering
\caption{The nonlinear mean correction $\bvarphi_{(2,0)}$ at the point $\triangle$. (Left) Axial (streamwise) velocity $u_{(2,0),x}$ and vertical velocity $u_{(2,0),y}=0$. (Middle) All four nonzero components of $\cG_{(2,0)} \in T_\mathbf{I}\mathrm{Pos}(3)$, the mean correction to the conformation tensor in its tangent form. (Right) All four nonzero components of $\C_{(2,0)} \in \mathrm{Pos}(3)$, the corresponding tensor from a standard expansion.}
\label{fig:EigenF20}
\end{figure}

\begin{figure}%[h]
% \psfrag{components of mathbfcalG22}[][][0.9]{components of $\cG_{(2,2)}$}
\includegraphics[width=\textwidth]{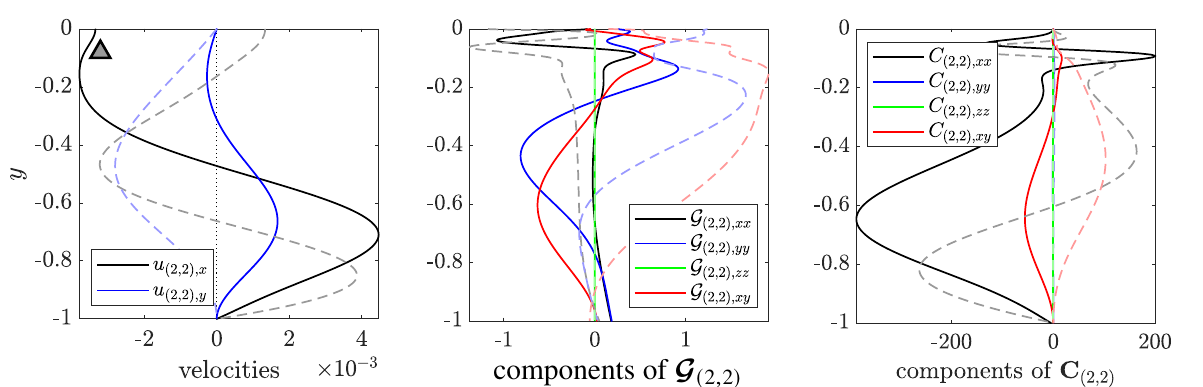}
\centering
\caption{Real (\sampleline{}) and imaginary (\sampleline{dashed}) parts of the nonlinear correction $\bvarphi_{(2,2)}$ at the point $\triangle$. (Left) Axial (streamwise) velocity $u_{(2,2),x}$ and vertical velocity $u_{(2,2),y}$. (Middle) All four nonzero components of $\cG_{(2,2)} \in T_\mathbf{I} \mathrm{Pos}(3)$. (Right) All four nonzero components of $\C_{(2,2)} \in \mathrm{Pos}(3)$, the corresponding tensor from a standard expansion. }
\label{fig:EigenF22}
\end{figure}

%
%  lower branch
%
We now examine the  bifurcation on the lower branch of the neutral curve for $Wi > 40$ to confirm the supercriticality predicted by the weakly nonlinear analysis. In Figure \ref{fig:compSUPERC}, bifurcation diagrams resulting from the weakly nonlinear analysis for 
the point $\triangle$ in \ Figure~\ref{fig:L500NC} are plotted with result from the  the Fourier-Chebyshev based branch continuation algorithm. The clear agreement we observe in the vicinity of the critical point confirms the existence of a stable supercritical state, and validates the weakly nonlinear predictions along the lower branch of the neutral curve. 

%
% fig 6
%
%
% fig 7
%

%
% fig 8
%

%
% 4.2
%
\subsection{Flow and polymer field prediction}
\label{sect:flowprediction}

The various flow and polymer fields generated as part of the weakly nonlinear analysis can be used to generate an approximation to the solution near to a bifurcation point.  The structure of the critical eigenfunction at the $\triangle$ in \ Figure~\ref{fig:L500NC} is shown in  Figure~\ref{fig:EigenF11}. The  flow and conformation tensor structures are  familiar from previous studies \citep{Garg2018,Khalid2021a} whereas the Cauchy-Green perturbation tensor $\cG_{(1,1)}$ hasn't been shown before. Figure~\ref{fig:EigenF11} shows that all components of $\cG_{(1,1)}$ are confined to the centerline of the channel.
For instance, ${\mathcal G}_{(1,1),xx}$ only develops a noticeable magnitude above $y=-0.2$.
On the other hand, $C_{(1,1),xx}$ indicates that the streamwise normal stretch reaches its maximum towards the bottom of the channel. This difference is explained by the shape of the laminar base state (cf.\ Figure~\ref{fig:base_state}).
$\C_b$ is smaller near the centerline, thus computing $ \cG_{(1,1)} = \F_b^{-1} \C_{(1,1)} \F_b^{-T}$ amplifies changes in that region, i.e., $ \cG_{(1,1)}$ recognizes deformations that are large relative to $\C_b$.
Again, this is an immediate consequence of the fact that the Riemannian metric on $\mathrm{Pos}(3)$ depends on the base point $\C_b$.
Physically, the new formulation highlights that the polymeric disturbance caused by the centre mode instability is confined to a small layer around the centerline, which would not be immediate from a standard expansion in $\C$ (cf.\ right panel of Figure~\ref{fig:EigenF11} or Figure 17 in \cite{Khalid2021a}).

\begin{figure}%[h]
\includegraphics[width=1.05\textwidth]{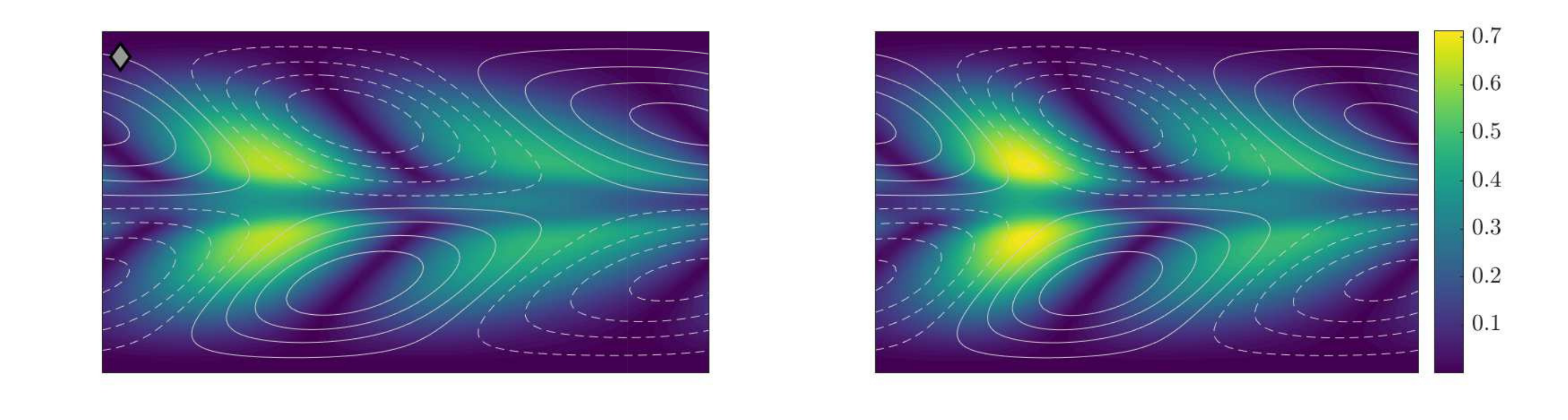}
\centering
\caption{Comparison of the supercritical state at point $\Diamond$ (identified in Figure~\ref{fig:compSUPERC}) as predicted by the weakly nonlinear analysis (left panel) and branch continuation (right panel) techniques. Contours show the geodesic distance between the base and full states $d(C_b,C) = \sqrt{ \mathrm{tr} \, \mathcal{G}^2}$, the lines correspond to the perturbation streamfunction.}
\label{fig:stateCOMP}
\end{figure}

Higher order disturbances are  more difficult to interpret on $\mathrm{Pos}(3)$, but up to $O(\varepsilon^2)$ can still be thought of as consecutive geodesic perturbations \citep{Hameduddin2019}. The $O(\varepsilon^2)$ terms from the weakly nonlinear expansion are given in Figure~\ref{fig:EigenF20}, which displays the first nonlinear mean correction $\bvarphi_{(2,0)}$, and Figure~\ref{fig:EigenF22}, which shows $\bvarphi_{(2,2)}$. With these fields known, the full flow state can be approximated by evaluating the weakly nonlinear expansion \eqref{eq:phi_expansion} up to second order, including $|A|$ in the shape functions as necessary. This low order approximation is compared with a full state from the continuation tool in Figure~\ref{fig:stateCOMP} at the point $\Diamond$ on the supercritical bifurcation branch (see Figure~\ref{fig:compSUPERC}).

%{\em Since the only natural way to measure deviations about an element of $\mathrm{Pos}(3)$ is along geodesics, one should really think of 
%$$
%\Vert \varepsilon \mathcal{G}_{(1,1)} \Vert_F = d \left(I,\exp \left( \varepsilon \mathcal{G}_{(1,1)} \right) \right) =  d \left(C_b,C_b + \varepsilon C_{(1,1)}\right) 
%$$
%as being the only correct notion of magnitude in assessing this initial disturbance.}
%\RK{Not sure where this sentence/definition fits into the narrative.. is the definition used?}
%\GB{-The only way it is used is through the eigenfunction figures: the 'equation' line means that perturbations should be measured as $\Vert \mathcal{G} \Vert$, so one should look at the $\mathcal{G}$ figures rather than the $C$ ones - I don't mind deleting this if you think it is unnecessary though-}

%
% Fig 9
%

%
% 4.3
%
\subsection{$\beta=0.9$, $L_{max}=100$ \& $Sc=10^6$}
\label{L=100}

\begin{figure}%[h]
\includegraphics[width=1\textwidth]{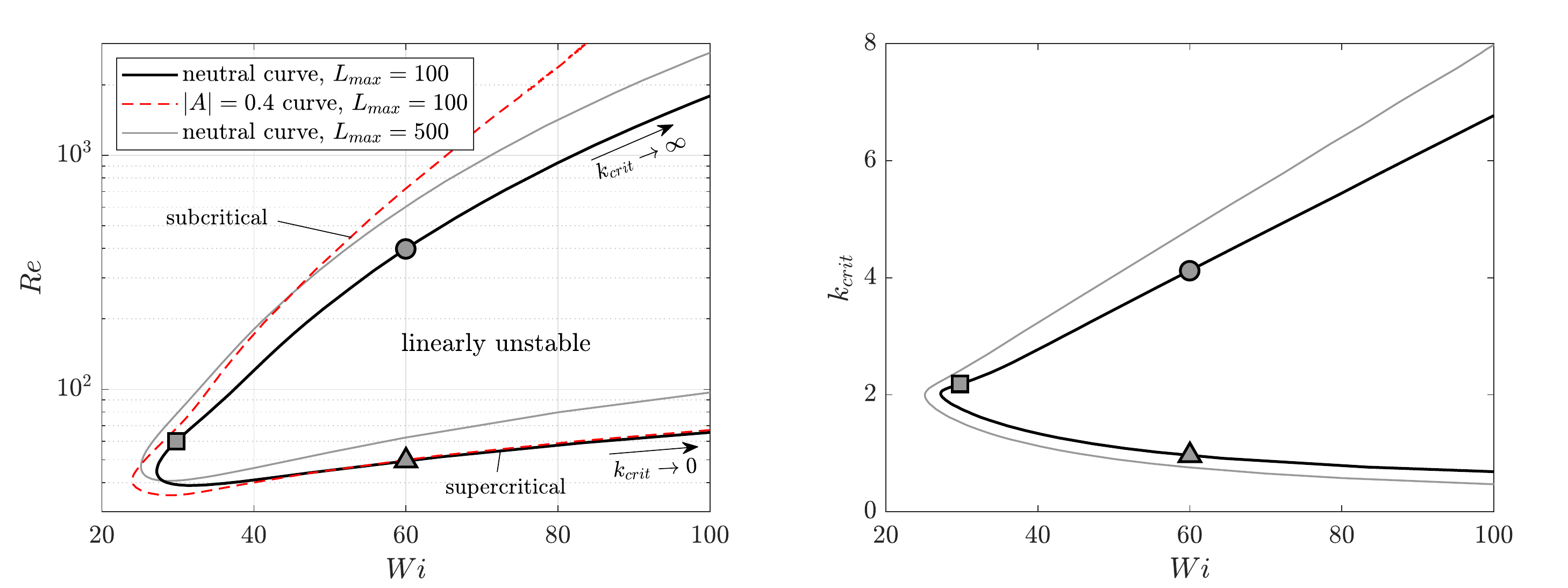}
\centering
\caption{(Left) Neutral curve corresponding to marginal linear stability at $\beta = 0.9$, $L_{max} = 100$, $Sc = 10^6$. Results of the weakly nonlinear analysis are shown in the form of a curve at steady state amplitude $\vert A \vert = 0.4$. The $L_{max} = 500$ neutral curve from Figure~\ref{fig:L500NC} is also shown for comparison. (Right) The development of the critical wave number, $k_{crit}$, along the neutral curve (corresponding curve for $L_{max}=500$ again shown in grey).}
\label{fig:neutralcurveL100}
\end{figure}

In this subsection, we  reduce $L_{max}$ to $100$ to explore less extensible (more realistic)  polymers and reintroduce the conformation tensor diffusion term into the governing equations \eqref{eq:fenep_C} by considering a finite Schmidt number, $Sc = 10^6$.  Figure \ref{fig:neutralcurveL100} shows the corresponding marginal stability curve complemented with a finite amplitude curve from the weakly nonlinear analysis.
The $L_{max} = 500$ neutral curve is also displayed for comparison in bright grey. All visible changes are caused by the adjustment of $L_{max}$: the introduction of finite $Sc$ alone has no visual effect. The key observation from  Figure \ref{fig:neutralcurveL100} is that reducing $L_{max}$ shifts the neutral curve down in $Re$, and reduces the slope of the lower branch.  In particular, lowering $L_{max}$ has a {\em destabilizing} effect in the elastic regime (low Reynolds numbers).  This counter-intuitive finding is the primary motivation for examining the $Re = 0$ instability recently found by \cite{Khalid2021b} at finite $L_{max}$ in \S\ref{sect:degenbeta}.

%
% Figure 10
%

\subsection{Energy analysis}
\label{sect:energy}

We now examine the energetic contributions of the different terms in the equations \eqref{eq:fenep} in order to examine the mechanisms driving the centre mode instability. This approach has proved useful to diagnose the character of instabilities - for example \cite{Joo1991,Joo1992} identified purely elastic instabilities in curved channel flows with this procedure (see also \cite{Zhang2013} and \cite{agarwal2014}).
Taking an $L^2$ inner product of the momentum equations at $O(\varepsilon)$ and the disturbance velocity field $\bu_{(1,1)} = (\bvarphi_{(1,1),1},\bvarphi_{(1,1),2})$ gives \citep[for more details see e.g.][]{Zhang2013} the disturbance kinetic energy equation
\begin{equation}
 \partial_t E := \frac12 \partial_t \Vert \bu_{(1,1)} \Vert_{L^2}^2 = \mathscr{P} + \mathscr{E} + \mathscr{W},
    \label{eq:kineticE}
\end{equation}
where 
$$
\mathscr{P} := - \frac12 \left\langle  \bnab \bu_b ,  \bu_{(1,1)} \otimes \bar{\bu}_{(1,1)} + \bar{\bu}_{(1,1)} \otimes \bu_{(1,1)} \right\rangle_{L^2}
$$
($\bar{\bu}$ is the complex conjugate of $\bu$) is the disturbance kinetic energy production due to the underlying shear of $\bu_b$,
$$
 \mathscr{E} := - \frac{\beta}{Re} \Vert \bnab \bu_{(1,1)} \Vert_{L^2}^2
$$
represents the viscous dissipation and is strictly negative, and
$$
\mathscr{W} := - \frac{(1-\beta)}{2Re} \left( \left\langle \bnab \bu_{(1,1)}, \T_{(1,1)} \right\rangle_{L^2} + \left\langle \T_{(1,1)}, \bnab \bu_{(1,1)} \right\rangle_{L^2} \right)
$$
indicates the rate of work done on the fluid by the polymeric stresses,
with
$$
\T_{(1,1)}:=D\T(\C_{b})\left[ \F_{b} \cG_{(1,1)} \F_{b}^{T} \right].
$$

Extending this procedure to identify the mechanisms behind the growth of elastic energy stored in the polymer is well known to be problematic \citep{Doering2006}. % \RK{Jacob...any relevant older refs?}). % not sure, this is fine
The underlying issue is that the elastic potential energy, which is a function of $\mathrm{tr} \, \C$, does not correspond to a norm in the obvious fashion that the kinetic energy does. 
Once again, this essentially comes down to the fact that the set $\mathrm{Pos}(3)$ does not constitute a linear vector space, 
and there is no notion of norm available.
% Therefore, the disturbances in $C$ should be measured along geodesics in $\mathrm{Pos}(3)$, according to the metric induced by the Riemannian structure.
This may be overcome by measuring disturbances in $\C$ along geodesics in $\mathrm{Pos}(3)$, according to the metric induced by the Riemannian structure. The work of \cite{Hameduddin2018} suggests that 
$$ 
\left(d(\C_b,\C)\right)^2 = \left(d({\bf I},\G)\right)^2 = \mathrm{tr} \left( \cG^H \cG \right),
$$ 
which immediately gives us a way of quantifying the evolution of polymer disturbances as
\begin{equation}
J:=\Vert d(\C_b,\C) \Vert_{L^2}^2 = \int_{[-1,1]} \mathrm{tr} \left( \cG^H (y) \cG (y) \right)   \, dy,
\label{eq:defJ}
\end{equation} % this is confusing b.c. it is innerprod over C, but this is the L^2 norm of mathcalG also.
% footnote some stuff about injectivity radius maybe, quote the nice thm
a formulation which was originally proposed in \cite{Hameduddin2019}. 
This, in fact, is the main advantage of relying on the alternative formulation of the governing equations given in equation \eqref{eq:fenep_G}. This newly defined quantity $J$ in \eqref{eq:defJ} is  equal to $\Vert \cG \Vert_{L^2}^2$ which is a natural  
%\RK{(true?)} 
%\GB{-yes, I guess you could interpret it as such - but this is a little backwards, because you do not define this quantity based on this similarity, you just end up with this after computing the natural generalization $\Vert d(C_b,C) \Vert_{L^2}$-}
generalization of the kinetic energy from \eqref{eq:kineticE}.
%modulo the algebraic manipulations from Section \ref{sect:C_expansion} and the fact that we are now considering $\mathcal{G}$ in the tangent space $T_I \mathrm{Pos}(3)$.
% This equivalence is due to the fact that the Riemannian metric on $\mathrm{Pos}(3)$ recovers the Frobenius inner product at the point $I$ on the manifold.
%This equivalence is immediate from the fact that the exponential map is locally an isometry by the Cartan-Hadamard theorem for connected complete manifolds with non-positive sectional curvatures ($\mathrm{Pos}(3)$ is a canonical example of such a manifold, see \cite{langfundamentals}), and that the Riemannian metric on $\mathrm{Pos}(3)$ recovers the Frobenius inner product at the point $I$ on the manifold. 

Adopting this polymer energy measure $J$, an energetic evolution equation for the polymer disturbances can now be obtained by taking an $L^2$ inner product of $\cG_{(1,1)}$ with the linearized disturbance equation (in a symmetric fashion) to obtain:
\begin{equation}
    \partial_t J =  \mathscr{A}_b + \mathscr{A}_1 + \mathscr{T} + \mathscr{E}_p,
    \label{eq:polyJ}
\end{equation}
where
% this uses the property tr(A^{T}B) = A_{ij}B_{ij}
% $$
% \mathscr{A} = \left\langle \mathcal{G}_{(1,1)}, 2 \mathrm{sym} \left( h\left( u_{(1,1)} \right) + \mathcal{G}_{(1,1)} h(u_{b}) \right) \right\rangle_{L^2} + \left\langle  2 \mathrm{sym} \left( h\left( u_{(1,1)} \right) + \mathcal{G}_{(1,1)} h(u_{b})\right)  ,\mathcal{G}_{(1,1)} \right\rangle_{L^2}
% $$
% captures the elastic energy growth induced by the velocity field, which is further subdivided into $\mathscr{A}_b$ and $\mathscr{A}_1$
$$
\mathscr{A}_b := \left\langle \cG_{(1,1)}, 2 \mathrm{sym} \left(  \cG_{(1,1)} h(\bu_{b}) \right) \right\rangle_{L^2} + \left\langle  2 \mathrm{sym} \left(  \cG_{(1,1)} h(\bu_{b}) \right),\cG_{(1,1)}  \right\rangle_{L^2}
$$
represents the contribution due to the base velocity field, %which is further subdivided into % do better.
$$
\mathscr{A}_1 := \left\langle \cG_{(1,1)}, 2 \mathrm{sym} \left( h\left( \bu_{(1,1)} \right)  \right) \right\rangle_{L^2} + \left\langle  2 \mathrm{sym} \left( h\left(\bu_{(1,1)} \right) \right) ,\cG_{(1,1)}  \right\rangle_{L^2}
$$
is the corresponding term capturing the effect of the disturbance velocity field $\bu_{(1,1)}$, % power transmitted from the disturbance velocity field into the polymer
$$
\mathscr{T} := -\left\langle \cG_{(1,1)}, \F_{b}^{-1} \T_{(1,1)} \F_{b}^{-T}  \right\rangle_{L^2}  -\left\langle \F_{b}^{-1} \T_{(1,1)} \F_{b}^{-T} , \cG_{(1,1)} \right\rangle_{L^2} 
$$
is the polymeric relaxation term and
$$
\mathscr{E}_p := \left\langle \cG_{(1,1)}, \frac{1}{ReSc} \F_{b}^{-1} \Delta \left( \F_{b} \cG_{(1,1)} \F_{b}^{T}\right)\F_{b}^{-T} \right\rangle_{L^2} + \left\langle \frac{1}{ReSc} \F_{b}^{-1} \Delta \left( \F_{b} \cG_{(1,1)} \F_{b}^{T}\right)\F_{b}^{-T} , \cG_{(1,1)} \right\rangle_{L^2}
$$
is the polymeric diffusion contribution.

\begin{figure}%[h]
% \psfrag{teveT}[][][0.8]{$\mathscr{T}$}
% \psfrag{teveA1}[][][0.8]{$\mathscr{A}_1$}
% \psfrag{teveAb}[][][0.8]{$\mathscr{A}_b$}
% \psfrag{teveE}[][][0.8]{$\mathscr{E}$}
% \psfrag{teveW}[][][0.8]{$\mathscr{W}$}
% \psfrag{teveP}[][][0.8]{$\mathscr{P}$}
\includegraphics[width=1\textwidth]{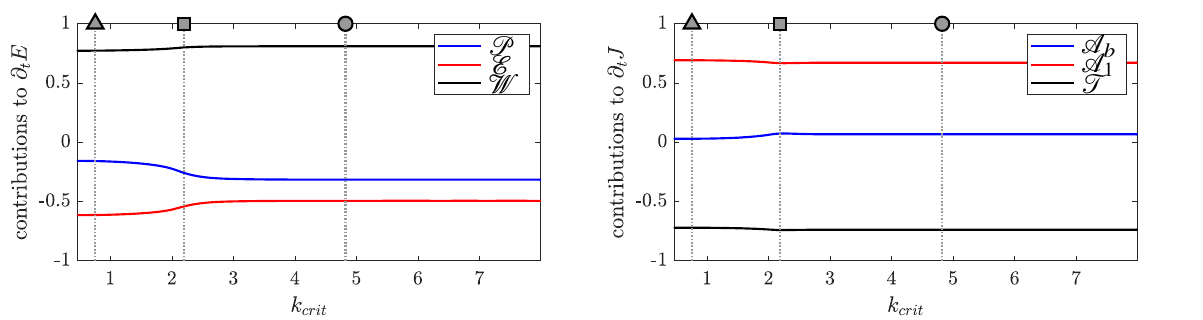}
\centering
\caption{Energy analysis results across  the $L_{max}=500$ neutral curve shown in Figure~\ref{fig:L500NC}. (Left) Components contributing to the production of the turbulent kinetic energy, $E$. (Right) Components contributing towards the evolution of the polymeric disturbance, $J$. All values are normalized across the neutral curve.}
\label{fig:energy}
\end{figure}

The contribution of each individual term along the neutral curve of subsection \ref{L=500} (parameterised by the wavenumber $k_{crit}$ which varies monotonically along the curve) is shown in Figure~\ref{fig:energy} for both the kinetic energy  equation \eqref{eq:kineticE} (left) and polymer `energy' equation \eqref{eq:polyJ}  (right). 
Based on the recent discovery of an inertialess linear instability that stems from the lower branch of the neutral curve \citep{Khalid2021b}, it was anticipated that the underlying destabilizing effects would be elastically driven along this branch. 
This is exactly what is seen: the polymer stress term is the sole energising term for the disturbance kinetic energy.
Figure~\ref{fig:energy}, however, indicates that this holds over the upper branch as well so that the centre-mode instability remains purely elastic - i.e., the rate of polymer work $\mathscr{W}$ is the only positive contribution to $\partial_t E$ - throughout the entirety of the neutral curve shown. Not even at $Re = 3000$ do we have a positive contribution from the turbulence production term, $\mathscr{P}$, which is the term that represents inertial effects and is responsible for the onset of instability in Newtonian turbulence.
In inertia-dominated flows, $\mathscr{P}$ is  the primary cause of turbulent kinetic energy growth \cite{Zhang2013}.

In the $J$ equation, the base flow ($\mathscr{A}_b$) barely contributes so that the effect of polymeric relaxation processes, $\mathscr{T}$, is balanced by the input of the perturbation velocity field through $\mathscr{A}_1$ (${\mathscr E}_P=0$ as $Sc \to \infty$ and so is not plotted).
The dominance of $\mathscr{A}_1$ which relies on the base polymer stretch rather than any  base flow shear confirms that the instability  mechanism is entirely elastic.

\begin{figure}%[h]
% \psfrag{teveT}[][][0.8]{$\mathscr{T}$}
% \psfrag{teveA1}[][][0.8]{$\mathscr{A}_1$}
% \psfrag{teveAb}[][][0.8]{$\mathscr{A}_b$}
% \psfrag{teveE}[][][0.8]{$\mathscr{E}$}
% \psfrag{teveW}[][][0.8]{$\mathscr{W}$}
% \psfrag{teveP}[][][0.8]{$\mathscr{P}$}
% \psfrag{teveEp}[][][0.8]{$\mathscr{E}_p$}
\includegraphics[width=1\textwidth]{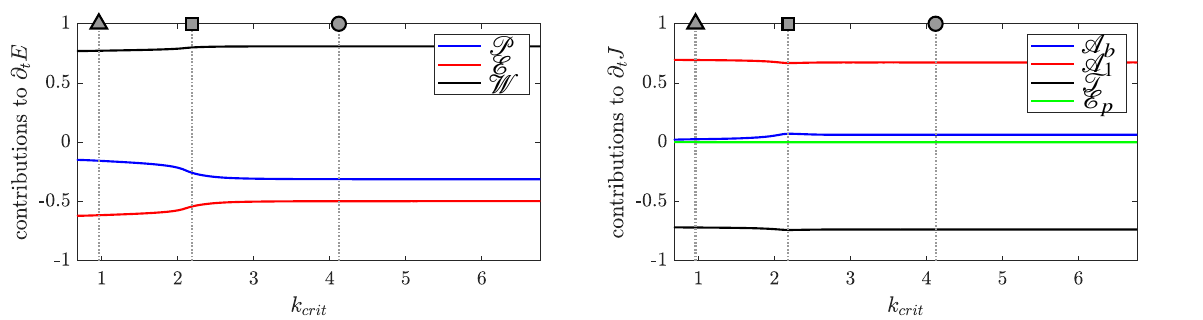}
\centering
\caption{Energy analysis results across  the $L_{max}=100$ neutral curve shown in Figure~\ref{fig:neutralcurveL100}. (Left) Components contributing to the production of the turbulent kinetic energy, $E$. (Right) Components contributing towards the evolution of the polymeric disturbance, $J$. All values are normalized across the neutral curve.}
\label{fig:energyL100}
\end{figure}

Choosing large but finite $Sc$ does not change this conclusion. Figure \ref{fig:energyL100} shows the energy analysis results for the neutral curve at $L_{max}=100$  in Figure \ref{fig:neutralcurveL100} of \S\ref{L=100}. Again, the polymeric viscous dissipation term, $\mathscr{E}_p$, does not contribute to the growth of $J$ (${\mathscr E}_p$ only starts to become significant for $Sc \sim O  ( 10^2)$)
%
%\RK{(check?)} \GB{-yes, correct-}) 
and the energy source for the instability is solely elastic.

%
% Fig 11
%

%
% Fig 12
%

%
%  4.5
%
\subsection{Inertialess limit}
\label{sect:degenbeta}

\begin{figure}%[h]
\includegraphics[width=1.025\textwidth]{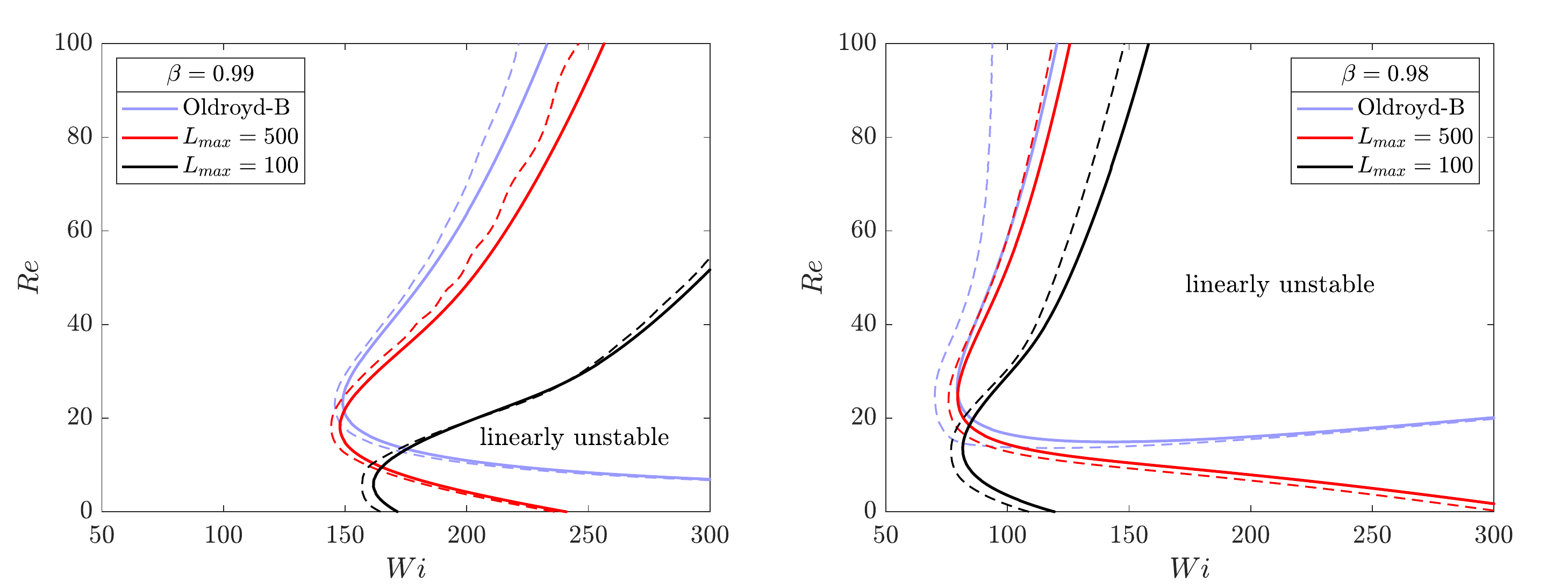}
\centering
\caption{Neutrally stable curves (\sampleline{}) around the inertialess ($Re=0$) limit for ultra-dilute polymer solutions at $\beta = 0.99$ (left) and $\beta = 0.98$ (right). The dashed lines (\sampleline{dashed}) are finite amplitude curves that show the nonlinear behaviour indicated by the weakly nonlinear analysis.}
\label{fig:degenbeta}
\end{figure}

In this section we explore the low-$Re$ elastic limit of the centre mode instability motivated by the finding in \S\ref{L=100} that decreasing $L_{max}$ makes the instability move to lower $Re$. Recent work \citep{Khalid2021b} has found the centre mode instability for $Re=0$ in the Oldroyd-B model, albeit at very high $Wi$ and very small $(1-\beta)$ i.e. the dilute limit. 
Our aim here is to see if we can find this instability at a lower, more realistic $Wi$ by varying $L_{max}$ in the FENE-P model. The effect of the viscosity ratio, $\beta$, for $L_{max} \to \infty$ (an Oldroyd-B fluid) is already known \citep[see inset (B) of Figure 2 in][]{Khalid2021b}. 
The instability first appears at  $\beta=0.9905$ with the critical $Wi$ decreasing as $\beta$ increases to 0.994, reaching a minimum of $Wi\approx 649$ (note their value $Wi'=973.8$ is defined using the base centreline speed)  
%
%\RK{(NB based on $U_{max}$...need to convert to our $Wi$)} \GB{-$Wi_c = 649 = 973.8*2/3$ is correct-} 
%
and then increases again as $\beta$ continues to increase beyond 0.994 towards 1. Thus, the lowest $\beta$ for which the $Re=0$ instability still exists (limited by the slope of the lower branch on the marginal curve) could also be decreased if the threshold $Wi$ for instability is decreased through adjusting $L_{max}$. This is what we find: see figure \ref{fig:degenbeta}, which shows that instability at $Re=0$ is possible at just over $Wi=100$ for $\beta=0.98$ and $L_{max}=100$. The finite amplitude curves generated by weakly nonlinear analysis and shown in figure~\ref{fig:degenbeta} further imply the existence of an unstable subcritical state in this inertialess regime. That is, the flow continues to be nonlinearly unstable when lowering $Wi$ below the threshold for linear instability.
 
%
% Fig 13
%

\begin{figure}%[h]
\includegraphics[width=1.025\textwidth]{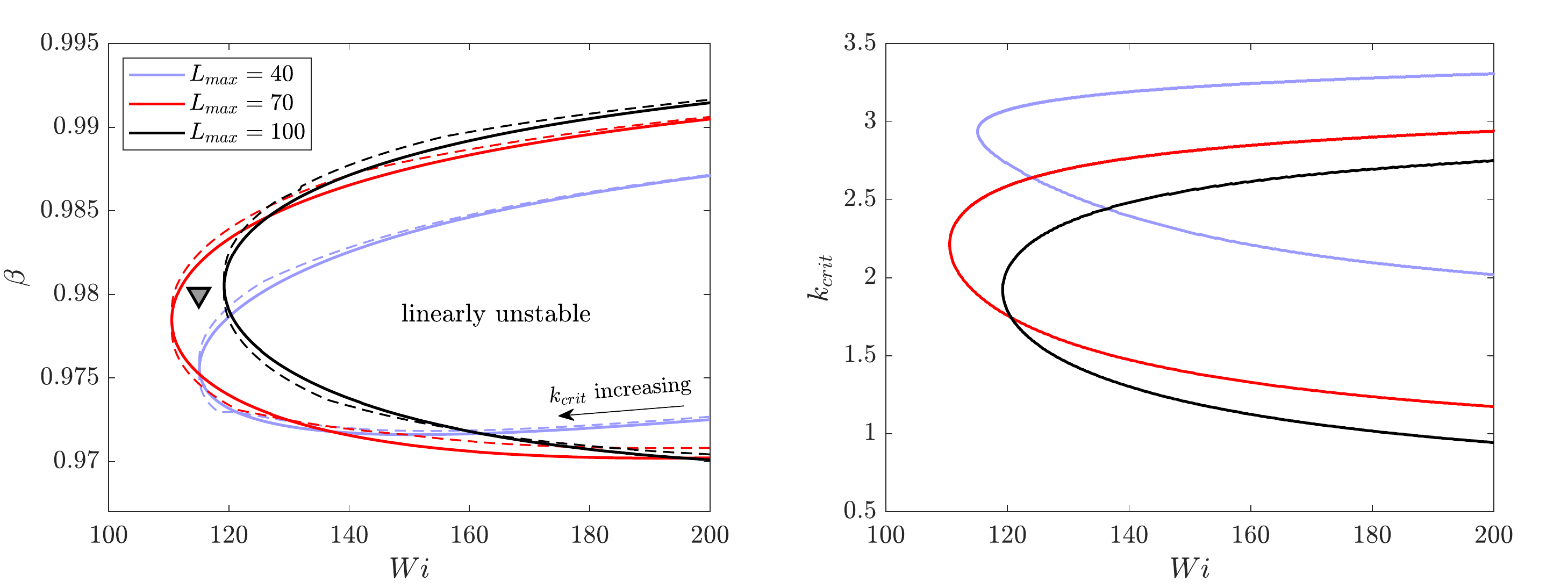}
\centering
\caption{(Left) Neutrally stable curves (\sampleline{}) at the inertialess limit $Re=0$ for ultra-dilute polymer solutions. The dashed lines (\sampleline{dashed}) are finite amplitude curves that show the nonlinear behaviour indicated by the weakly nonlinear analysis. (Right) Changes in the critical wave number, $k_{crit}$, as the neutral curves are traversed.}
\label{fig:Re0NC}
\end{figure}

Figure~\ref{fig:degenbeta} suggests further reduction in the threshold $Wi$ for instability may be possible by making  $L_{max}$ even smaller. Neutral curves in the $Wi-\beta$ plane at $Re=0$ for $L_{max}=40,70$ and $100$ are shown in  Figure~\ref{fig:Re0NC} along with the concomitant  finite amplitude curves.
Two important features are evident from this figure. Firstly, the destabilizing effect of $L_{max}$ has a limit, which appears to be in the interval $L_{max} \in [40,100]$ for $Re=0$.  Secondly, the weakly nonlinear analysis indicates that the bifurcation is subcritical with respect to $\beta$, except for high $Wi$ along the lower branch (in the $Wi-\beta$ plane) of the neutral curve where it becomes supercritical.

%
%  Fig 14
%

\begin{figure}%[h]
% \psfrag{dtJ}[][][0.7]{$ \quad \, \partial_t J$} 
% \psfrag{teveT}[][][0.7]{$\mathscr{T}$}
% \psfrag{teveA1}[][][0.7]{$\mathscr{A}_1$}
% \psfrag{teveAb}[][][0.7]{$\mathscr{A}_b$}
\includegraphics[width=0.5\textwidth]{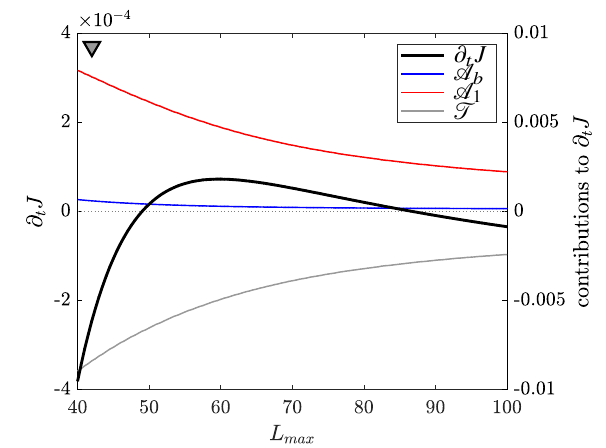}
\centering
\caption{The energy budget for the polymeric disturbance, $J$, at the inertialess limit, at point $\bigtriangledown$ ($Wi=115$, $\beta = 0.98$) in Figure \ref{fig:Re0NC}. Note that the scale for $\partial_t J$ (left axis) is enlarged to improve visibility.}
\label{fig:Re0J}
\end{figure}

The results of an energy budget analysis are shown in Figure~\ref{fig:Re0J} for this $Re=0$ instability at $Wi=115$ and $\beta=0.98$ - the $\bigtriangledown$ in Figure~\ref{fig:Re0NC} - as a function of $L_{max}$.  The kinetic energy evolution equation \eqref{eq:kineticE} is unable to handle the vanishing $Re$ situation and so  we exclusively focus on the budget in $J$, the measure introduced for polymeric perturbations.
Figure~\ref{fig:Re0J} tracks how the disturbance growth rate, $\partial_t J$, and each term contributing to it changes as $L_{max}$ is varied at point $\bigtriangledown$ (cf.\ Figure~\ref{fig:Re0NC}) ($\partial_t J=0$ indicate points on the neutral curve e.g. there is no instability at $L_{max}=100$ at $\bigtriangledown$). The contribution stemming from the base flow, $\mathscr{A}_b$, is still negligible, which indicates that stability is determined by the balance between (destabilizing) $\mathscr{A}_1$ and (stabilizing) $\mathscr{T}$.
The dissipation rate associated with polymeric relaxation processes, $\mathscr{T}$, becomes increasing negative as $L_{max}$ is decreased, ultimately causing stabilisation.  As expected from Figure~\ref{fig:Re0NC}, an optimal $L_{max}$ exists ($\approx 60$) for this particular pairing of $Wi$ and $\beta$. That it exists at all -  i.e. the FENE-P model is {\em more} unstable than the Oldroyd-B model  to this inertialess centre mode instability - is a surprise.

%The instability is caused by $\mathscr{A}_1$, which denotes the elastic contributions from the perturbation velocity field, $u_{(1,1)}$.
%$\mathscr{A}_1$ behaves similarly as $L_{max}$ is varied (albeit with a smaller base value and lower decay rate), although this dependence is not immediately obvious from the equations.
%The exact role of $L_{max}$ is better understood when considering a strictly positive Reynolds number.
%For $Re>0$, we know that the growth of $u_{(1,1)}$ is produced solely by  polymeric stresses through $\mathscr{W}$, which depends on $L_{max}$ in a similar fashion as $\mathscr{T}$, but is subject to different scaling.
%Thus, in the global context, there is a balance between the destabilizing effect of $\mathscr{W}$ and the stabilizing counterpart in $\mathscr{T}$.
%As $L_{max}$ is adjusted, there exists an optimal value for which the growth rate is maximal, depending on the scaling of these two terms.
%Of the two terms, only $\mathscr{W}$ is dependent on $Re$, which indicates that above a certain threshold in $Re$, $\mathscr{W}$ will be insignificant in comparison to $\mathscr{T}$, pushing the optimal extensibility to $L_{max} \to \infty$ (cf.\ Figure~\ref{fig:neutralcurveL100}).
%The only difference in the $Re=0$ case is that the effect of $\mathscr{W}$ manifests itself through $\mathscr{A}_1$ (compare this with the momentum equation), which is also apparent from Figure~\ref{fig:Re0J}.

%----------------------------------
%
% DISCUSSION
%
%----------------------------------

\section{Discussion}
\label{sect:conclusions}

% What has been done

% weakly nonlinear analysis
In this paper, we have considered the character of the bifurcation of a recently-discovered centre-mode  \citep{Garg2018,Khalid2021a} in rectilinear viscoelastic  channel flow for large $Re=O(10^3)$ down to the inertialess limit of  $Re=0$. Using weakly nonlinear analysis within a formal framework which respects the positive definiteness of the conformation tensor $\C$ \citep{Hameduddin2018,Hameduddin2019}, we find that the subcriticality found by \cite{Page2020} for one point of the neutral curve at $L_{max}=500$ is generic across the neutral curve and for  different $L_{max}$.
Supercriticality is only found at large $Wi$ on the `lower' (low-$Re$) branch of the neutral curve in the $(Wi,Re)$ plane otherwise the branch of travelling waves  arising from the neutral curve  reach down to lower  $Wi$ and the region where EIT is found. In this extended region of parameter space, the base  flow is nonlinearly unstable to disturbances of sufficient amplitude. The threshold amplitude to trigger this instability is determined by the minimal amplitude of approach of the stable manifold of the lower branch of travelling waves to the base flow. This is bounded from above by the amplitude of the lower branch itself and the one branch-tracking calculation done so far \citep[see Figure 3 in ][]{Page2020} indicates that this is small: the volume-averaged tr$\C$ of the travelling wave solutions stays within 5\% 
%
%\RK{(Jacob: a pure guess from  figure 3 of \cite{Page2020} - can you make more precise?)} % this is pretty accurate
%
 of the base flow value even when the $Wi$ is reduced to 50\% of its value at the bifurcation.  
%
%\RK{(meaning $Wi \approx 13$ is half of $26.9$ at the BP - clear?)}. %% JP I changed this as I'm not sure it was completely clear...  
%
Hence, for practical  purposes, the base flow may well appear linearly unstable below the neutral curve in $Wi$ (recent experiments suggest a similar situation in $Re$ \citep{Choueiri2021}). Assessing how far this situation continues as $Wi$ is decreased requires, of course, a full branch continuation procedure to map out the surface of travelling wave solutions.

% finite L_max
By using a FENE-P fluid we have also confirmed that the centre-mode instability persists  for  maximum polymer extension down to $L_{max}=40$ at least. Somewhat counterintuitively, the introduction of  finite $L_{max}$ is found to move the neutral curve closer to the inertialess $Re=0$ limit at fixed $\beta$. Pursuing this further by entering the dilute ($\beta \rightarrow 1$) limit, we also find that finite $L_{max}$ can bring the  linear instability recently found by \cite{Khalid2021b} down to more physically-relevant $Wi\gtrsim 110$ at $\beta=0.98$ compared with their threshold of $Wi\approx 649$ (based on the bulk velocity) 
at $\beta=0.994$ for $L_{max} \to \infty$. Again the instability is subcritical implying that inertialess rectilinear viscoelastic shear flow is nonlinearly unstable for even lower $Wi$. Assessing exactly how low again requires locating the saddle node (turning point) of the travelling waves as $Wi$ decreases which requires a branch continuation code.

% energy source
Finally, by considering the various energy terms in the disturbance kinetic energy equation, we have found  that the centre-mode instability is  purely elastic in origin even for $Re=O(10^3)$, rather than  `elasto-inertial', as the underlying shear does {\em not} energise the instability.  This finding is consistent with the recent smooth connection found by \cite{Khalid2021b} to an entirely elastic instability at $Re=0$ and suggests that EIT and ET may indeed be two different extremes of the same whole. Given that this instability is being suggested as the origin of EIT \citep{Garg2018,Page2020,Chaudhary2021, Khalid2021a}, the importance of inertia must emerge at finite amplitude and is perhaps already there in the travelling wave solutions especially when they establish their `arrowhead' form familiar from DNS at higher amplitudes and lower $Wi$ \citep{Dubief2020}.

% Experiments

In terms of experiments, the centre mode instability has recently been investigated in both channel \citep{Shnapp2021} and pipe flow \citep{Choueiri2021}. 
In a pipe, \citep{Choueiri2021} observed evidence of the centre mode instability at high $Wi=O(100)$ and low (subcritical) $Re$.
More relevant to the current results are the essentially inertialess ($Re\lesssim 0.3$) channel flow experiments of \citet{Shnapp2021}, which were 
conducted at very high $Wi \in (100,1700]$.
Finite amplitude traveling waves (or `elastic' waves in their terminology) were triggered by `small' disturbances -- in contrast to the `large' disturbances used in \cite{Pan2013} for $Re \lesssim 0.01$ and $Wi \lesssim 10$.  
Interestingly for the calculations performed here, they estimate the presence of a linear instability at $Wi=125 \pm 25$.
However, both studies were performed at considerably lower values of $\beta$ than those studied in the bulk of this paper ($\beta=0.74$ in \citet{Shnapp2021} and $\beta=0.56$ in \citet{Choueiri2021}). 
We have examined both of these solvent viscosities in Appendix \ref{sec:app_beta} and find that the significant reduction in $\beta$ leads to both (i) a smaller unstable region in the $Wi$-$Re$ plane and (ii) almost uniformly supercritical behaviour around the neutral curve.
This does not preclude the possibility that the branch may bend back down towards lower $Re$ and $Wi$, which cannot be captured in our third order weakly nonlinear analysis but which can be studied by branch continuation of the travelling waves.  

% the future

The obvious next steps after the analysis described here -- and particularly important in the context of the experimental observations at low $\beta$ -- is to employ a branch continuation procedure to track the travelling waves produced by the centre-mode instability to finite amplitudes and  then to explore where they exist in parameter space.
The inertialess limit is perhaps the most interesting but hardest to access numerically.  
These travelling waves, of course, provide their own launchpad for further (secondary) bifurcations from which subsequent solutions then suffer tertiary bifurcations and so forth. Establishing that this bifurcation cascade occurs precisely where EIT is observed in parameter space would provide convincing evidence of the  importance of the centre-mode instability. We hope to report on further progress in this direction in the near future.

\vspace{1cm}
\noindent
Acknowledgements: GB gratefully acknowledges the support of the Harding Foundation through a PhD scholarship
(https://www.hardingscholars.fund.cam.ac.uk).
%

%___________________________________________________________________________________________________________________________
%
% APPENDIX
%
%___________________________________________________________________________________________________________________________

\FloatBarrier
\appendix

\section{Numerical methods}
\label{sect:numerics}

The eigenvalue problem \eqref{eq:wnl_1} and the subsequent nonlinear equations in \eqref{eq:WNL} were solved using a Chebyshev discretization. Exploiting the symmetries of the centre eigenmode, the expansions were performed over half the channel width, $y \in [-1,0]$, with appropriate boundary conditions to enforce the symmetry of $u_x$, antisymmetry of $u_y$ and appropriate symmetries for the various components of $\C$. This approach crucially concentrates the  collocation points near {\em both} the channel boundary and the centreline where  the eigenmode is localised so that manageable truncations prove adequate.
For  the $\beta = 0.9$ neutral curves, $200$ Chebyshev were sufficient while higher $\beta$ values needed $300$-$400$ Chebyshev modes due to the increasing localization of the unstable eigenmode (see \cite{Khalid2021b}).
The neutral curves were obtained using a continuation scheme that relies on the tangent that the weakly nonlinear analysis yields.
Specifically, in the $Wi-Re$ plane, this is given by substituting $\vert A \vert=0$ into \eqref{eq:solvability}:
$$
\frac{Re_1}{Wi_1}=-\frac{\mathrm{Im}(\bar{d} b)}{\mathrm{Im}(\bar{d} a)}.
$$
In solving the eigenvalue problem, a shift-inverse spectral transformation \cite{MeerbergenKarl1994SaCt} was employed, using the eigenvalue at the previous continuation step, to isolate the critical eigenmode. The unstable mode was then obtained via standard power iteration. All results were cross-checked using two grid resolutions.

Results of the weakly nonlinear analysis were validated by an independently-developed branch continuation routine. In this, the flow solution is assumed to be steady in an appropriately-chosen Galilean frame (i.e. a travelling wave) which allows the time derivatives to be replaced by a spatial derivative in $x$ premultiplied by an {\em a priori} unknown phase speed $c=\omega_r/k$. The governing equations are then discretized in space using Fourier modes in $x$ and Chebyshev modes in $y$ across the domain $(x,y) \in [0,2\pi/k]\times [-1,1]$ to leave a high-dimensional - typically $O(10^5)$ degrees of freedom - nonlinear system of equations for the expansion coefficients. A good starting guess for the solution and $c$ can be generated near the  neutral curve and then the solver propagates along the solution surface via a pseudo-arclength continuation algorithm based on a Newton-Raphson iterative scheme (e.g. \cite{dijkstra_14}). 
Simulations for the curves appearing in Figures \ref{fig:compSUBC} and \ref{fig:compSUPERC} were run at $80$ Chebyshev and $40$ Fourier modes (so $7 \times 40 \times 80 \times 2=44,800$ real degrees of freedom). Resolution independence was carefully checked at the terminal point of each branch shown (using up to $80,000$ degrees of freedom). In this paper, the branch continuation code was only used to  confirm the weakly nonlinear analysis. A future report will describe it in detail when the results of using it to explore solution morphology a finite distance from the neutral curve will be presented.

\section{Results at moderate $\beta$}
\label{sec:app_beta}

\begin{figure}
\includegraphics[width=1.025\textwidth]{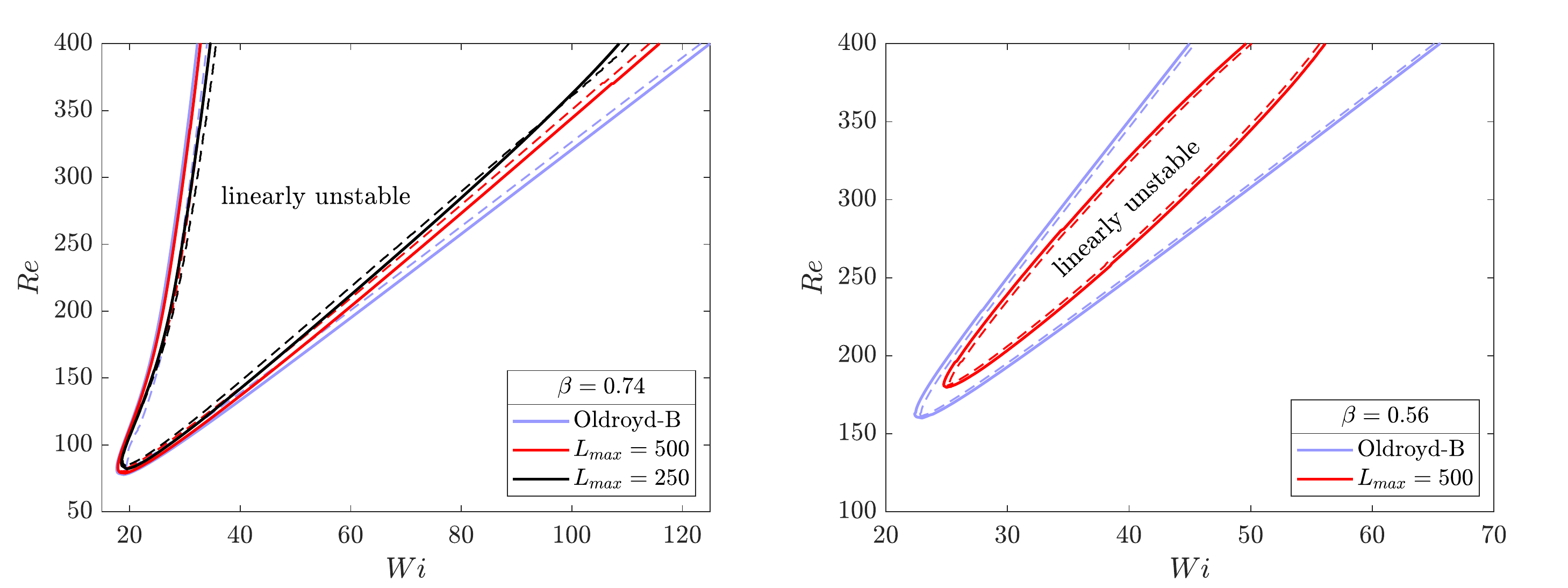}
\centering
\caption{
Neutrally stable curves (\sampleline{}) for low solvent viscosities $\beta = 0.74$ (left) and $\beta = 0.56$ (right). The dashed lines (\sampleline{dashed}) are finite amplitude curves that show the nonlinear behaviour indicated by the weakly nonlinear analysis.
}
\label{fig:low_beta}
\end{figure}
Motivated by recent experimental results \citep{Choueiri2021,Shnapp2021} at higher polymer concentrations, we briefly discuss the impact of reducing $\beta$ on both the linear instability and the predictions of our weakly nonlinear analysis. 
We consider two solvent viscosities, $\beta=0.74$ and $\beta=0.56$, which match the values obtained in \citet{Shnapp2021} and \citet{Choueiri2021} respectively (note the latter study was done in a pipe precluding any direct comparison here). 
Neutral curves and the weakly nonlinear results are reported in figure \ref{fig:low_beta} for both Oldroyd-B fluids and FENE-P fluid with relatively high $L_{max}$. 
The reduction in $\beta$ noticably shrinks the region of instability in the $Wi$-$Re$ plane, notably bending the lower part of the curve - which connects to $Re=0$ at high $\beta$ - upwards. 
Moreover, in contrast to the dilute ($\beta\geq 0.9$) results in the bulk of this paper, the introduction of finite extensibility has a uniformly stabilizing effect -- this behaviour is perhaps more typical of the more realistic polymer model; in many cases the reduction in the base-state normal stress tends to suppress more `interesting' Oldroyd-B results \citep[e.g. see the linear analyses in][]{Ray2014,Page2015}.

In addition, the weakly nonlinear results (dashed lines in figure \ref{fig:low_beta}) indicate almost uniformly supercritical behaviour around the neutral curve (note the small exception at high $Wi$ for $\beta=0.74$ and $L_{max}=250$). 
This finding should be contrasted to the recent experimental results at extreme $Wi\geq 100$ of \citet{Shnapp2021}, who have observed finite amplitude traveling waves at very low $Re$ at $\beta=0.74$, and motivates further study via branch continuation of exactly where nonlinear traveling waves are predicted to exist in the parameter space.

\bibliographystyle{jfm}

% \bibliographystyle{jfm}
% \begin{thebibliography}{}

% \bibitem[Page \& Kerswell (2019)]{Page2} \textsc{Page J.} and  \textsc{Kerswell R. R.} 2019 Koopman mode expansions between simple invariant solutions  \emph{J. Fluid Mech.} \textbf{879}, 1-27.

% \end{thebibliography}

\end{document}